\documentclass[a4paper,11pt,pdftex]{article}
\pdfoutput=1 
\usepackage{jheppub}

\usepackage{graphicx}
\usepackage{physics}
\usepackage{booktabs}

\makeatletter
\gdef\@fpheader{\ }
\makeatother

\newcommand{\lrpdv}[1]{\overset{\substack{\text{$\leftrightarrow$} \\ \vspace{-2.1ex}}}{\partial}{}^{#1}}

\title{
Tiny yet detectable WIMP-nucleon scattering cross sections in a pseudo-Nambu-Goldstone dark matter model
}

\author[a]{Tomohiro Abe,}
\author[a]{Kota Ichiki}
\affiliation[a]{
   Department of Physics and Astronomy, Faculty of Science and Technology, Tokyo University of Science,
   Yamazaki, Noda, Chiba 278-8510, Japan
}
\emailAdd{abe.tomohiro@rs.tus.ac.jp}
\emailAdd{6222504@ed.tus.ac.jp}

\abstract{
We investigate a pseudo-Nambu-Goldstone (pNG) dark matter (DM) model based on a gauged $SU(2)_x$ and a global $SU(2)_g$ symmetries. These symmetries are spontaneously broken to a global $U(1)_D$ symmetry by a vacuum expectation value of an $SU(2)_x \times SU(2)_g$ bi-fundamental scalar field. The global $SU(2)_g$ symmetry is also softly broken to a global $U(1)_D$ symmetry. Under the setup, a complex pNG boson arises. It is stabilized by $U(1)_D$ and is a DM candidate. Its scattering cross section off a nucleon is highly suppressed by small momentum transfer and thus evades the stringent constraints from DM direct detection experiments. Assuming all the couplings in the dark sector are real, a discrete symmetry arises. Consequently, in addition to the pNG DM, the lighter one of an $SU(2)_x$ gauge boson $V^0$ and a CP-odd scalar boson $a_0$ from the bi-fundamental scalar field can also serve as a DM candidate. Therefore, the model provides two-component DM scenarios. 
We find that the relic abundance of the DM candidates explains the measured value of the DM energy density. 
We also find that the pNG DM is the dominant DM component in large regions of the parameter space. 
In contrast to the pNG DM, both $V^0$ and $a_0$ scatter off a nucleon, and their scattering cross sections are not suppressed.
However, their scattering event rates are suppressed by their number densities. 
Thus, the scattering cross section is effectively reduced. 
We show that the effective WIMP-nucleon scattering cross sections in the two-component scenarios are smaller than the current upper bounds and above the neutrino fog.
}

\begin{document}

\maketitle

\section{Introduction}

There is overwhelming observational evidence for dark matter (DM). 
A popular scenario explaining the measured value of the DM energy density, 
$\Omega_\text{DM} h^2 = 0.120 \pm 0.001$~\cite{1807.06209}, 
is the thermal DM scenario. In the scenario, DM particles are in the thermal bath consisting of the standard model (SM) particles in the early universe and decouple as the universe expands. The DM energy density is explained by the freeze-out mechanism~\cite{Lee:1977ua}.  
It requires annihilation processes of DM particles into particles in the thermal bath.  
Therefore, the thermal DM interacts with the SM particles, and elastic scattering processes of DM off a SM particle are expected.

DM direct detection experiments, such as the XENONnT~\cite{XENON:2023cxc}, the PandaX-4T~\cite{PandaX:2024qfu}, and the LUX-ZEPLIN (LZ)~\cite{LZCollaboration:2024lux} experiments, aim to detect DM-nucleus scattering events. No significant signal has been detected, and stringent upper bounds on the DM-nucleon scattering cross section have been established, placing tight constraints on thermal DM models. Consequently, the DM-nucleon scattering amplitude must be highly suppressed.

Pseudo-Nambu-Goldstone (pNG) DM models~\cite{Gross:2017dan, Balkin:2018tma, Alanne:2018zjm, Karamitros:2019ewv, Jiang:2019soj, 2001.03954,2001.05910, Abe:2021nih, Abe:2021byq, Okada:2021qmi, Cai:2021evx, Abe:2022mlc, Liu:2022evb, Otsuka:2022zdy, Abe:2024vxz, Biswas:2024wbz} naturally evade the constraints from the direct detection experiments while keeping the annihilation cross section for the freeze-out. In those models, the scattering amplitude involving the pNG DM is proportional to the squared momentum transfer $q^2$ from the dark sector to the visible sector. 
This transfer is on the order of the DM mass, $q^2 \simeq m_\text{DM}^2$, in the DM annihilation processes, but takes a much smaller value in DM-nucleon scattering in the direct detection experiments, $\abs{q} = \mathcal{O}(10)~\text{MeV}$.
As a result, the pNG DM models predict highly suppressed DM-nucleon scattering cross section $\sigma_\text{SI}$ while retaining a canonical annihilation cross section value, $\expval{\sigma v} \simeq 2 \times 10^{-26}$~cm$^3$ s$^{-1}$.

In the simplest pNG DM model~\cite{Gross:2017dan}, pNG DM arises from the spontaneous symmetry breaking of a global $U(1)$ symmetry in the dark sector.
A complex scalar $S$, charged under this global $U(1)$ symmetry, acquires a vacuum expectation value (VEV), and the global $U(1)$ symmetry is spontaneously broken.
As a result, a Nambu-Goldstone (NG) boson $\chi$ arises.
The global $U(1)$ symmetry is also explicitly broken to $Z_2$ by a mass-dimension-two operator $S^2$, which generates mass of $\chi$.
Under this $Z_2$ symmetry, $S$ transforms as $S \to -S$.
This $Z_2$ symmetry is also spontaneously broken by the VEV of $S$. Consequently, the model suffers from the domain-wall problem.

It is possible to avoid the domain-wall problem by embedding the spontaneously broken discrete symmetry into a gauge symmetry. 
Models based on a $U(1)$ gauge symmetry in the dark sector~\cite{2001.03954,2001.05910, Liu:2022evb} successfully avoid the domain-wall problem. However, the models predict decaying dark matter due to the kinetic mixing between the $U(1)$ gauge symmetry and the SM hypercharge gauge symmetry $U(1)_Y$. To make the DM sufficiently long-lived, the breaking scale of the newly introduced gauge symmetry has to be higher than $\mathcal{O}(10^{10})$~GeV.
An alternative model without requiring a much higher energy scale than the DM mass is proposed in~\cite{Abe:2022mlc}. 
The model is based on $U(1)_\text{gauge} \times SU(2)_\text{global} \to U(1)_\text{global}$, where $SU(2)_\text{global}$ is softly broken into $U(1)_\text{global}$. In the model, the DM is a stable complex scalar boson protected by the $U(1)_\text{global}$ symmetry.
However, a large value of the $U(1)_\text{gauge}$ gauge coupling is required to obtain the measured value of the DM energy density by the freeze-out mechanism. Thus, the gauge coupling blows up at high energy due to the renormalization group evolution. It is resolved by embedding the $U(1)_\text{gauge}$ symmetry into an $SU(2)$ gauge symmetry~\cite{Otsuka:2022zdy}. 
In another model based on $U(1)_\text{gauge} \times U(1)_\text{global}$~\cite{Abe:2024vxz}, the $U(1)_\text{gauge}$ gauge coupling also blows up at high energy.

In this paper, we propose a new pNG DM model based on $SU(2)_x \times SU(2)_g$, 
where $SU(2)_x$ is a gauge symmetry, and $SU(2)_g$ is a global symmetry that is softly broken to a global $U(1)_g$ symmetry. A bi-fundamental scalar field $\Phi$ is introduced, and its VEV spontaneously breaks $SU(2)_x \times SU(2)_g$ to a global $U(1)_D$ symmetry. This spontaneous symmetry breaking generates five NG bosons. Three of them are eaten by the gauge bosons associated with $SU(2)_x$. The remaining two NG bosons acquire mass from the soft breaking of the global $SU(2)_g$ symmetry and are pNG bosons. Since the two pNG bosons have the same mass due to $U(1)_D$, they are represented by a single complex scalar field $\chi$. Thanks to the global $U(1)_D$ symmetry, $\chi$ can be stable and a DM candidate.
Unlike other pNG models with a $U(1)$ gauge symmetry, our model does not predict the Landau pole of the gauge coupling because $SU(2)$ is asymptotic free.
Our model contains only one VEV in the dark sector in contrast to the pNG model proposed in \cite{Otsuka:2022zdy}, which is based on an $SU(2)$ gauge symmetry with two VEVs.
As we will discuss below, some couplings in the scalar potential are generally complex. 
For simplicity, we assume all the couplings are real. 
This assumption enhances the symmetry in the model and leads to two-component DM scenarios.
As a result, the model has a rich phenomenology and predicts WIMP-nucleon scattering cross sections that are smaller than the current upper bound but above the neutrino fog. 

This paper is organized as follows.
In section~\ref{sec:model}, we describe our model. 
The model predicts different DM scenarios depending on the mass spectra. We focus on the following three cases:  
(i) a single-component DM scenario in which $\chi$ is the DM candidate,
(ii) a two-component DM scenario in which both $\chi$ and an $SU(2)_x$ gauge boson $V^0$ are the DM candidates,
and 
(iii) a two-component DM scenario in which both $\chi$ and $a_0$, which is a CP-odd scalar in $\Phi$, are the DM candidates.
In section~\ref{sec:constraints}, we discuss constraints on the model from the SM-Higgs couplings to the SM particles, the Higgs invisible decay, and the perturbative unitarity bounds on the scalar quartic and gauge couplings. 
In section~\ref{sec:single-component-DM-scenario}, a single-component DM scenario is discussed.
In section~\ref{sec:chi-V-scenario}, we discuss a two-component DM scenario in which $\chi$ and $V^0$ are the DM candidates.
In section~\ref{sec:chi-a-scenario}, we discuss the other two-component DM scenario in which $\chi$ and $a_0$ are the candidates for DM.
Section~\ref{sec:conclusion} is devoted to our conclusion.
In appendix~\ref{app:some-identities}, we discuss some identities for constructing the scalar potential. 
In appendix~\ref{app:custodial}, we discuss an accidental global $SU(2)$ symmetry that guarantees the degenerate mass of the $SU(2)_x$ gauge bosons.
In appendix~\ref{sec:enhanced-symmetry}, we discuss the symmetry structure in the scalar potential in the model.

\section{Model}\label{sec:model}
We impose an $SU(2)_x$ gauge symmetry and an approximate global $SU(2)_g$ symmetry in the dark sector. We assume the global $SU(2)_g$ symmetry is explicitly broken to a global $U(1)_g$ symmetry by a soft-breaking term.
All the SM fields are singlet under the $SU(2)_x \times SU(2)_g$ symmetry. 
We add a scalar field $\Phi$ that is a bi-fundamental representation of $SU(2)_x \times SU(2)_g$ and is a singlet under the SM gauge symmetry. 
The VEV of $\Phi$ breaks $SU(2)_x \times SU(2)_g$ to a global $U(1)_D$ symmetry. 
This spontaneous symmetry breaking would generate five NG bosons if $SU(2)_g$ were exact symmetry. 
Because we assume $SU(2)_g$ is softly broken into $U(1)_g$, 
the true symmetry breaking pattern is $SU(2)_x \times U(1)_g \to U(1)_D$, and thus only three NG bosons arise, and they are eaten by the $SU(2)_x$ gauge bosons.
Therefore, after the spontaneous symmetry breaking, we have two pNG bosons. 
Since they have the same mass due to $U(1)_D$, we express them by a single complex scalar field $\chi$. Thanks to the global $U(1)_D$ symmetry, $\chi$ can be stable and is a dark matter candidate.

\subsection{The fields in the dark sector}
The bi-fundamental scalar field $\Phi$ transforms under the symmetry as
\begin{align}
 \Phi \to U_x \Phi U_g^\dagger,
\end{align}
where $U_x$ and $U_g$ are two-by-two unitary matrices for $SU(2)_x$ and $SU(2)_g$, respectively.
In the following, we introduce a soft-symmetry breaking term that explicitly breaks $SU(2)_g$ to $U(1)_g$. Then, $\Phi$ transforms as
\begin{align}
 \Phi \to U_x \Phi e^{-i \frac{\tau^3}{2} \theta_g},
\end{align}
where $\tau^3$ is a Pauli matrix.
We assume $\Phi$ develops the following VEV,
\begin{align}
 \expval{\Phi} = 
\frac{1}{\sqrt{2}}
\begin{pmatrix}
 0 & 0 \\ 0 & v_s
\label{eq:desired-vacuum}
\end{pmatrix}
.
\end{align}
This VEV induces a spontaneous symmetry breaking, $SU(2)_x \times SU(2)_g \to U(1)_D$. Here, $U(1)_D$ is a diagonal subgroup of $SU(2)_x \times SU(2)_g$, and $\Phi$ transforms under $U(1)_D$ as
\begin{align}
 \Phi \to e^{i \frac{\tau^3}{2} \theta_D} \Phi e^{-i \frac{\tau^3}{2} \theta_D}.
\label{eq:U(1)D-and-Phi}
\end{align}
This symmetry breaking generates five NG bosons.
We parametrize $\Phi$ as
\begin{align}
 \Phi  
=
\begin{pmatrix}
 \frac{\sigma_1 + i a_0}{\sqrt{2}} & i \pi_{V^+} \\
 i \bar{\chi} & \frac{v_s + \sigma_2 - i \pi_{V^0}}{\sqrt{2}}
\end{pmatrix}
,
\label{eq:Phi-components}
\end{align}
where $\sigma_1$, $\sigma_2$, $a_0$, and $\pi_{V^0}$ are real scalar fields, $\bar{\chi}$ and $\pi_{V^+}$ are complex scalar fields, and $\bar{\chi}= \chi^* $. 
Among the scalar fields, $\pi_{V^0}$, $\pi_{V^+}$, and $\pi_{V^-} = (\pi_{V^+})^*$ are would-be NG bosons eaten by the $SU(2)_x$ gauge fields, $V^0$, $V^+$, and $V^-$.
There are two more NG bosons, $\chi$ and $\bar{\chi}$, associated with $SU(2)_x \times SU(2)_g \to U(1)_D$. 
The explicit breaking $SU(2)_g \to U(1)_g$ generates the mass of $\chi$ and $\bar{\chi}$ as we will see below. 
From eqs.~\eqref{eq:U(1)D-and-Phi} and \eqref{eq:Phi-components}, we find 
the $U(1)_D$ charge of each field. It is summarized in table~\ref{tab:U1D-charge}.
\begin{table}[tbp]
 \centering
\begin{tabular}{c|c|c|c|c|c}
\toprule
 field         & $V^\pm$ & $\pi_{V^\pm}$ & $\chi$  & $\bar{\chi}$ & {all other fields}\\ \hline
 $U(1)_D$ charge  & $\pm 1$ & $\pm 1$       & $+1$    & $-1$             & $0$             \\ 
\bottomrule 
\end{tabular}
\caption{The $U(1)_D$ charge of each field.
Here, $V^\pm$ are the $SU(2)_x$ gauge fields given in eq.~\eqref{eq:def-of-Vmu},
$\pi_{V^{\pm}}$ are would-be NG bosons and eaten by $V^\pm$, 
$\chi$ is the pNG DM, and $\bar{\chi}$ is the anti-particle of $\chi$.
}
\label{tab:U1D-charge}
\end{table}
This global $U(1)_D$ symmetry is unbroken, and thus, the lightest $U(1)_D$ charged particle is stable and a DM candidate.

For later convenience, we introduced $\varphi_1$ and $\varphi_2$ as
\begin{align}
\varphi_1 = \mqty( i \chi \\  \frac{\sigma_1 - i a_0}{\sqrt{2}} ),
\quad 
\varphi_2 = \mqty( i \pi_{V^+} \\ \frac{v_s + \sigma_2 - i \pi_{V^0}}{\sqrt{2}} ).
\label{eq:varphi_12}
\end{align}
Using these fields, $\Phi$ is written as
\begin{align}
 \Phi = 
\begin{pmatrix}
 \tilde{\varphi}_1 & \varphi_2
\end{pmatrix}
,
\end{align}
where $\tilde{\varphi}_1 = \epsilon \varphi_1^*$, and 
\begin{align}
 \epsilon = 
\begin{pmatrix}
 0 & 1 \\ -1 & 0
\end{pmatrix}
.
\end{align}
Under the $SU(2)_x \times U(1)_g$ symmetry, $\varphi_1$ and $\varphi_2$ transform as
\begin{align}
 \varphi_1 \to& U_x e^{i\frac{1}{2} \theta_g} \varphi_1,\\
 \varphi_2 \to& U_x e^{i\frac{1}{2} \theta_g} \varphi_2.
\end{align}

\subsection{Scalar potential}
The scalar potential of this model consists of two parts. 
One is symmetric under the $SU(2)_x \times SU(2)_g$.
The other softly breaks $SU(2)_g$ to $U(1)_g$.
We denote them $V_0$ and $V_\text{soft}$, respectively.
They are given by\footnote{
We could introduce $\sum_{a = 1}^3 \kappa^a \tr(\Phi^\dagger \phi \tau^a)$
as $V_\text{soft}$. However, it is transformed into the term in Eq.~\eqref{eq:Vsoft} 
by redefining $\Phi$ without lose of generality~\cite{Abe:2022mlc, Otsuka:2022zdy}.
Other operators allowed by the gauge symmetry are reduced into the terms in eq.~\eqref{eq:V0}, 
see appendix~\ref{app:some-identities}.
}
\begin{align}
 V_0
=&
\mu_H^2 H^\dagger H
+
\mu_1^2 \tr(\Phi^\dagger \Phi)
+ 
\left(
\mu_2^2 \det(\Phi)
+ (h.c.)
\right)
\nonumber\\
&
+ \frac{\lambda_H}{2}  (H^\dagger H)^2
+ \tilde{\lambda}_{1} H^\dagger H \tr(\Phi^\dagger \Phi) 
+ 
\left(
\tilde{\lambda}_{2} H^\dagger H \det(\Phi)
+ (h.c.)
\right)
\nonumber\\
&
+ \frac{\lambda_1}{2} \left(\tr(\Phi^\dagger \Phi)\right)^2
+ \left( \lambda_2 \tr(\Phi^\dagger \Phi) \det(\Phi) + (h.c.) \right)
\nonumber\\
&
+ \left( \lambda_3 (\det(\Phi))^2 + (h.c.) \right)
+ 2\lambda_4 \abs{\det(\Phi)}^2
,
\label{eq:V0}
\\
 V_{\text{soft}}
=& 
\frac{1}{2}\mu^2_\chi \tr(\Phi^\dagger \Phi \tau_3),
\label{eq:Vsoft}
\end{align}
where $H$ is the SM Higgs field,
\begin{align}
 H = \begin{pmatrix}
      i \pi_{W^+} \\ \frac{v + \sigma_3 - i \pi_Z}{\sqrt{2}}
     \end{pmatrix}.
\end{align}
Here, $v$ is related to the Fermi constant $G_F$ as $v = \qty(\sqrt{2} G_F)^{-1/2} \simeq 246$~GeV.
We can also express the scalar potential by using $\varphi_1$ and $\varphi_2$ as
\begin{align}
V_0 + V_\text{soft}
=&
\mu_H^2 H^\dagger H
+
\mu_1^2 
\left(
\varphi_1^\dagger \varphi_1  + \varphi_2^\dagger \varphi_2
\right)
+
\left(
 \mu_2^2 \varphi_1^\dagger \varphi_2  
+(h.c.)
\right)
\nonumber\\
&
+ \frac{\lambda_H}{2}  (H^\dagger H)^2
+ \tilde{\lambda}_{1} H^\dagger H 
\left(
\varphi_1^\dagger \varphi_1  + \varphi_2^\dagger \varphi_2
\right)
+ 
\left(
 \tilde{\lambda}_{2} H^\dagger H 
\varphi_1^\dagger \varphi_2  
+ (h.c.)
\right)
\nonumber\\
&
+ \frac{\lambda_1}{2}  \left( \varphi_1^\dagger \varphi_1  + \varphi_2^\dagger \varphi_2 \right)^2
+ \left( \lambda_2 \left( \varphi_1^\dagger \varphi_1  + \varphi_2^\dagger \varphi_2 \right)
 \varphi_1^\dagger \varphi_2 + (h.c.)\right) 
\nonumber\\
&
+
\left(
\lambda_3  \left(\varphi_1^\dagger \varphi_2 \right)^2  + (h.c.) 
\right)
+
2 \lambda_4 (\varphi_1^\dagger \varphi_2)(\varphi_2^\dagger \varphi_1)
\nonumber\\
& + \frac{1}{2}\mu_\chi^2 \left( \varphi_1^\dagger \varphi_1 - \varphi_2^\dagger \varphi_2  \right).
\label{eq:potential}
\end{align}

\subsection{Kinetic term}
The kinetic term of the scalar field is given by
\begin{align}
 \tr(D^\mu \Phi^\dagger D_\mu \Phi)
= D^\mu \varphi_1^\dagger D_\mu \varphi_1 + D^\mu \varphi_2^\dagger D_\mu \varphi_2,
\end{align}
where
\begin{align}
 D_\mu \Phi  =& \partial_\mu \Phi + i g_D V_\mu \Phi,\\
 D_\mu \varphi_1 =&  \partial_\mu \varphi_1 + i g_D V_\mu \varphi_1,\\
 D_\mu \varphi_2 =&  \partial_\mu \varphi_2 + i g_D V_\mu \varphi_2,
\end{align}
with the $SU(2)_x$ gauge coupling $g_D$, and the $SU(2)_x$ gauge fields, 
\begin{align}
 V_\mu  = \mqty( \frac{1}{2} V^0_\mu & \frac{1}{\sqrt{2}} V^+_\mu \\ \frac{1}{\sqrt{2}} V^-_\mu & - \frac{1}{2} V^0_\mu ).
\label{eq:def-of-Vmu}
\end{align}

\subsection{Discrete symmetry \texorpdfstring{$C_\text{dark}$}{}}\label{subsec:Z2}

The scalar potential contains four complex parameters, $\mu_2^2$, $\tilde{\lambda}_2$, $\lambda_2$, and $\lambda_3$.
If all the parameters in the scalar potential are real, the scalar potential is invariant under a transformation, $\Phi \to \Phi^*$. The kinetic terms of $\Phi$ and $V^\mu$ are also invariant under this transformation if the gauge fields transform as $V^\pm \to -V^\mp$ and $V^0 \to -V^0$. Therefore, 
if $\mu_2^2$, $\tilde{\lambda}_2$, $\lambda_2$ and $\lambda_3$ are real, 
the Lagrangian is invariant under the following field transformation:
\begin{align}
 \chi \to& - \bar{\chi},\label{eq:Cdark_chi}\\ 
 a_0 \to& - a_{0},\\ 
 V^\pm \to& - V^{\mp},\quad \pi_{V^{\pm}} \to - \pi_{V^{\mp}},\\ 
 V^0 \to& - V^{0},\quad  \pi_{V^{0}} \to - \pi_{V^{0}}. \label{eq:Cdark_pi} 
\end{align}
All the SM fields, $\sigma_{1}$, and $\sigma_2$ do not transform under this discrete symmetry.
This discrete symmetry is a $Z_2$ symmetry and is regarded as charge conjugation in the dark sector.
We refer to this discrete symmetry as $C_\text{dark}$.

In this paper, we assume all the parameters in the scalar potential are real.
This assumption reduces the number of the free parameters in the model and makes analysis easy.
Also, it induces $C_\text{dark}$.
This $C_\text{dark}$ symmetry can stabilize $a_0$ or $V^0$ in addition to the lightest $U(1)_D$ charged particle.
Thus, the model contains two-component DM scenarios. 
Depending on the mass spectra, different particles are DM candidates. We discuss it in section~\ref{sec:DM-candidate}.

\subsection{Scalar mass eigenstates}
From the stationary condition of the scalar potential, we find
\begin{align}
 \mu_H^2 =& - \frac{\tilde{\lambda}_1 v_s^2 + \lambda_H v^2}{2},\\
 \mu_1^2 =& - \frac{\lambda_1 v_s^2 + \tilde{\lambda}_1 v^2 - \mu_\chi^2}{2},\\
 \mu_2^2 =& - \frac{\tilde{\lambda}_2 v^2 + \lambda_2 v_s^2 }{2}.
\end{align}
The mass terms of the scalar fields are given by
\begin{align}
 {\cal L}_{mass}
=&
- m_\chi^2 \bar{\chi} \chi
- \frac{m_{a_0}^2}{2} a_0^2
- \frac{1}{2}
\begin{pmatrix}
 \sigma_1 & \sigma_2 & \sigma_3
\end{pmatrix}
\mathcal{M}^2
\begin{pmatrix}
 \sigma_1 \\ \sigma_2 \\ \sigma_3
\end{pmatrix}
,
\end{align}
where
\begin{align}
 m_\chi^2 =& \mu_\chi^2,\\
 m_{a_0}^2 =& \mu_\chi^2 + (-\lambda_3 + \lambda_4) v_s^2, \label{eq:ma2}\\
\mathcal{M}^2=&\begin{pmatrix}
 (\lambda_3 +\lambda_4) v_s^2 + \mu_\chi^2 & \lambda_2 v_s^2 & \tilde{\lambda}_2 v v_s \\
 \lambda_2 v_s^2 & \lambda_1 v_s^2 & \tilde{\lambda}_1 v v_s \\
  \tilde{\lambda}_2 v v_s & \tilde{\lambda}_1 v v_s & \lambda_H v^2 
\end{pmatrix}
\label{eq:CP-even-mass-mat}
.
\end{align}
We define the mass eigenstates for the CP-even scalars $(h_1, h_2, h_3)$ by
\begin{align}
 \begin{pmatrix}
 \sigma_1 \\ \sigma_2 \\ \sigma_3
\end{pmatrix}
=
R
\begin{pmatrix}
 h_1 \\ h_2 \\ h_3
\end{pmatrix}
,
\end{align}
where $R$ is an orthogonal matrix and satisfies
\begin{align}
R^t \mathcal{M}^2 R
=
diag.(m_{h_1}^2, m_{h_2}^2, m_{h_3}^2)
.
\label{eq:RM2R=diag}
\end{align}
Here, $m_{h_1}$, $m_{h_2}$, and $m_{h_3}$ are the mass of $h_1$, $h_2$, and $h_3$, respectively.
We identify $h_1$ with the SM-like Higgs boson, and $m_{h_1} \simeq 125$~GeV.
We parametrize each component of $R$ as
\begin{align}
R =&
 \begin{pmatrix}
 R_{11} & R_{12} & R_{13} \\ 
 R_{21} & R_{22} & R_{23} \\ 
 R_{31} & R_{32} & R_{33} 
 \end{pmatrix}
\nonumber\\
=& 
\mqty(0 & \cos\theta_3 & \sin\theta_3 \\ 0 & -\sin\theta_3 & \cos\theta_3 \\ 1 & 0 & 0)
\mqty(\cos\theta_h & 0 & -\sin\theta_h \\ 0 & 1 & 0 \\ \sin\theta_h & 0 & \cos\theta_h)
\mqty(1 & 0 & 0 \\ 0 & \cos\theta_1 & \sin\theta_1 \\ 0 & -\sin\theta_1 & \cos\theta_1)
\nonumber\\
=&
\mqty(
\sin\theta_h \sin\theta_3 
& \cos\theta_1 \cos\theta_3 - \cos\theta_h \sin\theta_1 \sin\theta_3 
& \cos\theta_3 \sin\theta_1 + \cos\theta_h \cos\theta_1  \sin\theta_3 
\\
\sin\theta_h \cos\theta_3 
& -\cos\theta_1 \sin\theta_3 - \cos\theta_h \sin\theta_1 \cos\theta_3 
& -\sin\theta_3 \sin\theta_1 + \cos\theta_h \cos\theta_1 \cos\theta_3 
\\
\cos\theta_h 
& \sin\theta_h \sin\theta_1 
& -\sin\theta_h \cos\theta_1 
)
.
\end{align}
From eqs.~\eqref{eq:ma2}, \eqref{eq:CP-even-mass-mat}, and \eqref{eq:RM2R=diag},
the quartic couplings are expressed by $R_{ij}$ and the masses of the scalar bosons as
\begin{align}
 \lambda_H=& \sum_{j=1}^3 \frac{m_{h_j}^2 R_{3j}^2}{v^2},\label{eq:lamH}\\
 \lambda_1=& \sum_{j=1}^3 \frac{m_{h_j}^2 R_{2j}^2}{v_s^2},\\
 \tilde{\lambda}_1=& \sum_{j=1}^3 \frac{m_{h_j}^2 R_{3j}R_{2j}}{v v_s},\\
 \lambda_2=& \sum_{j=1}^3 \frac{m_{h_j}^2 R_{1j}R_{2j}}{v_s^2},\\
 \tilde{\lambda}_2=& \sum_{j=1}^3 \frac{m_{h_j}^2 R_{3j}R_{1j}}{v v_s},\\
 \lambda_3 =& -\frac{m_{a_0}^2}{2 v_s^2} + \sum_{j=1}^3 \frac{m_{h_j}^2 R_{1j}^2}{2 v_s^2},\label{eq:lam3}\\
 \lambda_4 =& \frac{m_{a_0}^2 - 2 m_\chi^2}{2 v_s^2} + \sum_{j=1}^3 \frac{m_{h_j}^2 R_{1j}^2}{2 v_s^2}.\label{eq:lam4}
\end{align}

\subsection{Gauge sector}\label{sec:gauge-sector}
The mass terms of the $SU(2)_x$ gauge bosons are obtained from the kinetic term of $\Phi$. We find
\begin{align}
\mathcal{L}_\text{gauge mass}
=&
  m_{V^+}^2 V^-_\mu V^{+\mu}
+ \frac{m_{V^0}^2}{2} V^0_\mu V^{0 \mu} 
,
\end{align}
where
\begin{align}
 m_{V^+} = m_{V^0} =& \frac{g_D v_s}{2} \equiv m_V. \label{eq:gD}
\end{align}
The degeneracy in $m_{V^+}$ and $m_{V^0}$ is due to an accidental global $SU(2)$ symmetry among $(\pi_{V^+}, \pi_{V^-}, \pi_{V^0})$, similar to the custodial symmetry in the SM. However, it is explicitly broken by the mass difference between $m_\chi$ and $m_{a_0}$. 
The detail is discussed in appendix~\ref{app:custodial}.
We find that the mass difference between $m_{V^+}$ and $m_{V^0}$ is induced at the loop level and is $\mathcal{O}(1)$~GeV. 
Such a small mass difference does not change the following results.
In the following analysis, we do not include the loop correction and use the tree level relation $m_{V} = m_{V^+} = m_{V^0}$.

\subsection{Model parameters}
As shown in
eqs.~\eqref{eq:lamH}--\eqref{eq:lam4} and \eqref{eq:gD}, 
the parameters in the scalar potential and the $SU(2)_x$ gauge coupling are expressed by 
the following 11 parameters:
\begin{align}
 \qty(v,\ v_s,\ m_{h_1},\ m_{h_2},\ m_{h_3},\ m_\chi,\ m_{a_0},\ m_{V},\ \theta_h,\ \theta_2,\ \theta_3).
\end{align}
Two of them are already measured, $v \simeq 246$~GeV, $m_{h_1} \simeq 125$~GeV. 
Thus, we have nine free parameters in the model. 
In the following analysis, we determined $v_s$ to obtain the measured value of the DM energy density,
$\Omega_\text{DM} h^2 = 0.120 \pm 0.001$~\cite{1807.06209}, by the freeze-out mechanism.

\subsection{DM candidates}
\label{sec:DM-candidate}

There are two global symmetries: the global $U(1)_D$ symmetry and the discrete symmetry $C_\text{dark}$ discussed in Sec.~\ref{subsec:Z2}.
They stabilize some particles in the dark sector and generate DM candidates in the model. 

The global $U(1)_D$ symmetry is summarized in table~\ref{tab:U1D-charge}.
The lightest $U(1)_D$-charged particle is stabilized. Namely, the lighter of $V^+$ and $\chi$ is a DM candidate. 
The corresponding anti-particle is also a DM candidate.
In this paper, we focus on the pNG DM and assume $V^+$ is unstable.
There are two-body decay channels, $V^+ \to \chi h_j$ for $m_{V^+} \geq m_\chi + m_{h_j}$. If this decay mode is kinematically forbidden, $V^+$ decays via three-body decay channels with off-shell $h_j$, such as $V^+ \to \chi h_j^* \to \chi \gamma \gamma $. Therefore, $V^+$ decays for $m_{V^+} > m_\chi$.
In the following analysis, we assume $m_V > m_\chi$. Therefore, $\chi$ is always a DM candidate thanks to the global $U(1)_D$ symmetry.

As discussed in section~\ref{subsec:Z2}, we assume all the parameters in the scalar potential are real, and the Lagrangian is invariant under the discrete symmetry $C_\text{dark}$.
The $C_\text{dark}$ symmetry restricts decay channels of $V^0$ and $a_0$.
There are following decay modes for $V^0$ and $a_0$, 
\begin{align}
V^0 \to& \chi \bar{\chi},\ a_0 h_j^{(*)},\\   
a_0 \to& \chi V^{-(*)},\ \bar{\chi} V^{+(*)},\ V^0 h_j^{(*)}.
\end{align}
Note that decay processes of $V^0$ and $a_0$ into SM particles, such as $V^0 \to \text{SM SM}$, are forbidden by $C_\text{dark}$,  
and thus one of $V^0$ and $a_0$ can be stable.
If $m_{a_0} > m_{V^0}$, then $a_0 \to V^0 h_j^{(*)}$ is always possible, and thus $a_0$ is unstable.
In that case, $V^0 \to \chi \bar{\chi}$ is the only possible decay mode of $V^0$. 
For $m_{V^0} < 2 m_\chi$, it is kinematically forbidden, and $V^0$ is a DM candidate in addition to $\chi$. As a result, the model is classified into a two-component DM model.
Similarly, if $m_{a_0} < m_{V^0}$, $V^0$ can always decay via $V^0 \to a_0 h_j^{(*)}$, but $a_0$ can decay only if 
$a_0 \to \chi V^{-(*)}$ is kinematically allowed. Since $m_V > m_\chi$, it is possible for $m_{a_0} > 2 m_\chi$. Thus, $a_0$ is stable for $m_{a_0} < 2 m_\chi$ if $m_{a_0} < m_V$.
Therefore, if the mass of the lighter one of $V^0$ and $a_0$ is smaller than $2m_\chi$, it is stable and a DM candidate in addition to $\chi$. 

To summarize, the number of DM candidates depends on the mass spectra. 
As we discussed above, we assume $\chi$ is always stable in the following analysis and study the following three cases.
\begin{description}
 \item[Case 1] For $2 m_\chi < m_{V^0}$ and $2 m_\chi < m_{a_0}$, $\chi$ (and its anti-particle $\bar{\chi}$) is only the DM candidate in the model, and thus the model is classified into a single-component DM scenario. 
 \item[Case 2] For $m_{V^0} < 2 m_\chi$ and $m_{V^0} < m_{a_0}$, both $\chi$ and $V^0$ are stable, and the model is classified into a two-component DM scenario.
 \item[Case 3] For $m_{a_0} < 2 m_\chi$ and $m_{a_0} < m_{V^0}$, both $\chi$ and $a_0$ are stable, and a two-component DM scenario is realized.
\end{description}

\subsection{Some couplings}
The $\chi\chi h_j$, $a_0 a_0 h_j$, $V^0 V^0 h$, and $ffh$ couplings play an important role in the processes of the DM direct detection, 
where $f$ stands for the SM fermions. 
They are given by
\begin{align}
 {\cal L} \supset&
 -\sum_{j=1}^3 g_{\chi \chi h_j} \chi \bar{\chi} h_j
 -\frac{1}{2} \sum_{j=1}^3 g_{aa h_j} a_0^2 h_j
 - \sum_{j=1}^3 \sum_{f} g_{ffh} \bar{f}f h_j
\nonumber\\
&
+ \sum_{j=1}^3 \frac{1}{2} g_{V^0V^0 h_j} V^0_\mu V^{0\mu} h_j
,
\end{align}
where
\begin{align}
 g_{\chi \chi h_j} =& \frac{m_{h_j}^2 R_{2j}}{v_s},\label{eq:g_h-chi-chi}\\
 g_{aa h_j} =& \frac{m_{h_j}^2+2 (m_{a_0}^2 - m_\chi^2)}{v_s} R_{2j}, \label{eq:g_h-a-a} \\
 g_{ffh_j} =& \frac{m_f}{v} R_{3j}, \label{eq:g_h-f-f} \\
 g_{V^0 V^0 h_j} =& \frac{2 m_V^2}{v_s} R_{2j}.
\end{align}
The following three-point interaction terms are important to calculate the relic abundance of $V^0$ and $a_0$.
\begin{align}
\mathcal{L}
\supset&
-i g_{V a_0 \chi} V_{\mu}^+ \left( a_0 \lrpdv{\mu} \bar{\chi} \right)
+i g_{V a_0 \chi} V_{\mu}^- \left( a_0 \lrpdv{\mu} \chi \right)
\nonumber\\
&
+ \sum_{j=1}^3 g_{V h_j \chi} V_{\mu}^0 \left( h_j \lrpdv{\mu} a_0 \right)
- i g_{V \chi \chi} V_{\mu}^0 \left( \bar{\chi} \lrpdv{\mu} \chi \right)
,
\end{align}
where
\begin{align}
 g_{V h_j \chi} =  g_{V h_j a_0} =& \frac{m_V}{v_s} R_{1j},\\
 g_{V a_0 \chi}  = g_{V \chi \chi}  =& \frac{m_V}{v_s}.
\end{align}

\section{Constraints}\label{sec:constraints}

\subsection{Higgs coupling}

The SM-like Higgs couplings to the SM fermions and the SM electroweak gauge bosons ($W$ and $Z$) are given by 
\begin{align}
 \mathcal{L} 
\supset&
 - \sum_{f} g_{ffh} \bar{f}f h_1
 - g_{WWh_1} W^{+ \mu} W^-_{\mu} h_1
 - \frac{1}{2} g_{ZZh_1} Z^{\mu} Z_{\mu} h_1,
\end{align}
where
\begin{align}
 g_{ffh_1} =& \frac{m_f}{v} \cos\theta_h = g_{ffh}^\text{SM} \cos\theta_h,\\
 g_{WWh_1} =& \frac{2m_W^2}{v} \cos\theta_h = g_{WWh}^\text{SM} \cos\theta_h,\\
 g_{ZZh_1} =& \frac{2 m_Z^2}{v} \cos\theta_h = g_{ZZh}^\text{SM} \cos\theta_h.
\end{align}
They are given by the SM prediction times $\cos\theta_h$.
These couplings are precisely measured in the ATLAS~\cite{ATLAS:2022vkf} and CMS~\cite{CMS:2022dwd} experiments.
The ATLAS experiment found
\begin{align}
 \kappa_V = 1.035 \pm 0.031,\\
 \kappa_f = 0.95 \pm 0.05,
\end{align}
with a positive correlation, 39\%.
The CMS experiment found\footnote{
The 95\% CL contours are given in HEPdata \url{https://www.hepdata.net/record/ins2104672}.
}
\begin{align}
 \kappa_V = 1.014,\\
 \kappa_f = 0.906.
\end{align}
In our model, $\kappa_V = \kappa_f = \cos\theta_h$ at the tree level, and we find
\begin{align}
\begin{cases}
 \abs{\sin\theta_h} < 0.238  & \text{ (ATLAS)} \\
 \abs{\sin\theta_h} < 0.23   & \text{ (CMS)}
\end{cases}
, 
\end{align}
at 95\% CL.

\subsection{Higgs invisible decay}

The SM-like Higgs boson can decay into a pair of pNG DM particles for $m_{h_1} > 2 m_\chi$.
The partial decay width is given by
\begin{align}
 \Gamma(h_1 \to \chi \bar{\chi})
= \frac{1}{8\pi} \frac{1}{2m_\chi} g_{\chi \chi h_1}^2 \sqrt{1 - \frac{4 m_\chi^2}{m_{h_1}^2}},
\end{align}
where $g_{\chi\chi h_1}$ is given in~eq.~\eqref{eq:g_h-chi-chi}.
This decay mode is searched in the ATLAS and the CMS experiments, and they found the upper bounds on the Higgs invisible decay branching ratio as 
\begin{align}
 \text{BR}_\text{inv} <  
\begin{cases}
 0.107 & \text{(ATLAS \cite{ATLAS:2023tkt})} \\
 0.15 & \text{(CMS \cite{CMS:2023sdw})}
\end{cases}
,
\end{align}
at 95\% CL. This upper bound gives a strong constraint on the model for $m_{h_1} > 2 m_\chi$.

\subsection{Perturbative unitarity}
We estimate the upper bound on the scalar quartic couplings and the gauge coupling from the perturbative unitarity (PU) bound~\cite{Lee:1977eg}. 
We calculate two-to-two scattering amplitudes $\mathcal{M}_{fi}$ in high energy limit, where $i$ and $f$ indicate the initial and final states.
The amplitude is decomposed into partial waves as
\begin{align}
 \mathcal{M}_{fi} = 16 \sum_\ell (2\ell + 1) P_\ell(\cos\theta) a^\ell_{fi},
\end{align}
where $P_\ell$ is the Legendre polynomial, and $\cos\theta$ is a scattering angle.
The PU bound is given by
\begin{align}
 \abs{\text{Re} \lambda^0} < \frac{1}{2},
\end{align}
where $\lambda^0$ is any eigenvalue of the matrix $a^0_{fi}$.
We use this condition to estimate the upper bounds on the scalar quartic and gauge couplings.

\subsubsection{Scalar couplings}
We consider two-to-two scattering processes in which all the particles in the initial and final states are scalar bosons.
There are 12 real scalar fields, including would-be NG bosons. There are 78 ways to construct two-particle states from them, and thus $a^0_{fi}$ is a $78\times78$ matrix. Most of the eigenvalues are zero, and some eigenvalues are degenerated. We find nine non-zero and independent eigenvalues resulting in the following PU bound,
\begin{align}
 \abs{-2\lambda_3 + \lambda_1} <& 8\pi,\label{eq:pu-scalar-0}\\
 \abs{\pm\tilde{\lambda}_2 + \tilde{\lambda}_1} <& 8\pi,\\
 \abs{-6 \lambda_3 + \lambda_1 + 4 \lambda_4} <& 8\pi,\\
 \abs{ \lambda_1 - 2 \lambda_4 } <& 8\pi,\\
 \abs{ \lambda_1 + \lambda_3 + \lambda_4 \pm \sqrt{4 \lambda_2^2 + (\lambda_3 - \lambda_4)^2} } <& 8\pi,\\
 \abs{ \lambda_H } <& 8\pi,\\
 \abs{ x_i } <& 16\pi,\label{eq:pu-scalar-last}
\end{align}
where $x_i$ are three solutions of the following equation,
\begin{align}
0= & x^3 -6(\lambda_H  + 2 \lambda_1 + 2 \lambda_3 + 2 \lambda_4 )x^2
\nonumber\\
&
+
\Biggl(
-144 \lambda_2^2 - 32 \tilde{\lambda}_2^2 + 
  120 \lambda_3 \lambda_1 + 20 \lambda_1^2 - 32 \tilde{\lambda}_1^2 + 
  48 \lambda_3 \lambda_4 \nonumber\\
& \qquad \qquad
+ 88 \lambda_1 \lambda_4 + 32 \lambda_4^2 
+ 72 \lambda_H \qty(\lambda_1 + \lambda_3 + 72 \lambda_4)
\Biggr)
x 
\nonumber\\
&
+
8 \Biggl(
-96 \lambda_2 \tilde{\lambda}_1 \tilde{\lambda}_2
+8 \left(5 \lambda_1+2 \lambda_4\right) \tilde{\lambda}_2^2
+6 \lambda_3 
   \left(
     -15 \lambda_1 \lambda_H
     -6 \lambda_4 \lambda_H
     +8 \tilde{\lambda}_1^2
    \right)
\nonumber\\
& \qquad \qquad
+8 \lambda_1 \tilde{\lambda}_1^2
+32 \lambda_4 \tilde{\lambda}_1^2
-15 \lambda_1^2 \lambda_H
+108 \lambda_2^2 \lambda_H
-24  \lambda_4^2 \lambda_H
-66 \lambda_1 \lambda_4 \lambda_H
\Biggr).
\end{align}
We impose eqs.~\eqref{eq:pu-scalar-0}--\eqref{eq:pu-scalar-last} in the following analysis.

\subsubsection{Gauge coupling}
We can also estimate the upper bound on $g_D$ from the PU bound.
Diagrams exchanging a gauge boson in $t$-channel or $u$-channel are logarithmically divergent in high energy limit. In that case, we cannot utilize the PU bound to obtain the upper bound on couplings.
To avoid the logarithmic divergence, we investigate $V^0 \chi \to V^0 \chi$, where $V^0$ is transversely polarized. We find its scattering amplitude as
\begin{align}
 \mathcal{M}
=& - \frac{g_D^2}{4}(1 + \lambda \lambda'),
\end{align}
where $\lambda$ and $\lambda'$, which can be $\pm 1$, are helicities of the gauge bosons in the initial and final states, respectively. As a result, the perturbative unitarity bound on $g_D$ is given by
\begin{align}
 g_D < \sqrt{16\pi}.
\end{align}

\section{Single-component DM scenario}\label{sec:single-component-DM-scenario}
For $m_{V^0} > 2 m_\chi $ and $m_{a_0} > 2 m_\chi$, only the pNG DM $\chi$ and its anti-particle $\bar{\chi}$ are the DM candidates. Thus, the model realizes a single-component DM scenario.
In this section, we study this scenario.

\subsection{Relic abundance}
\begin{figure}[tbp]
  \centering
  \includegraphics[width=0.195\hsize]{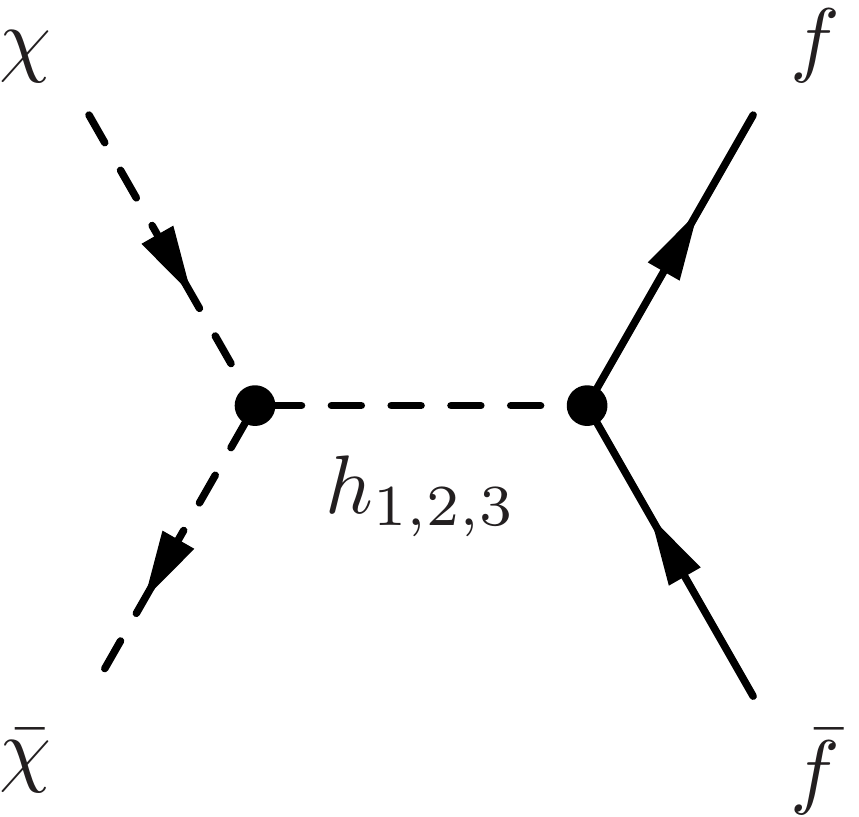}\qquad
  \includegraphics[width=0.23\hsize]{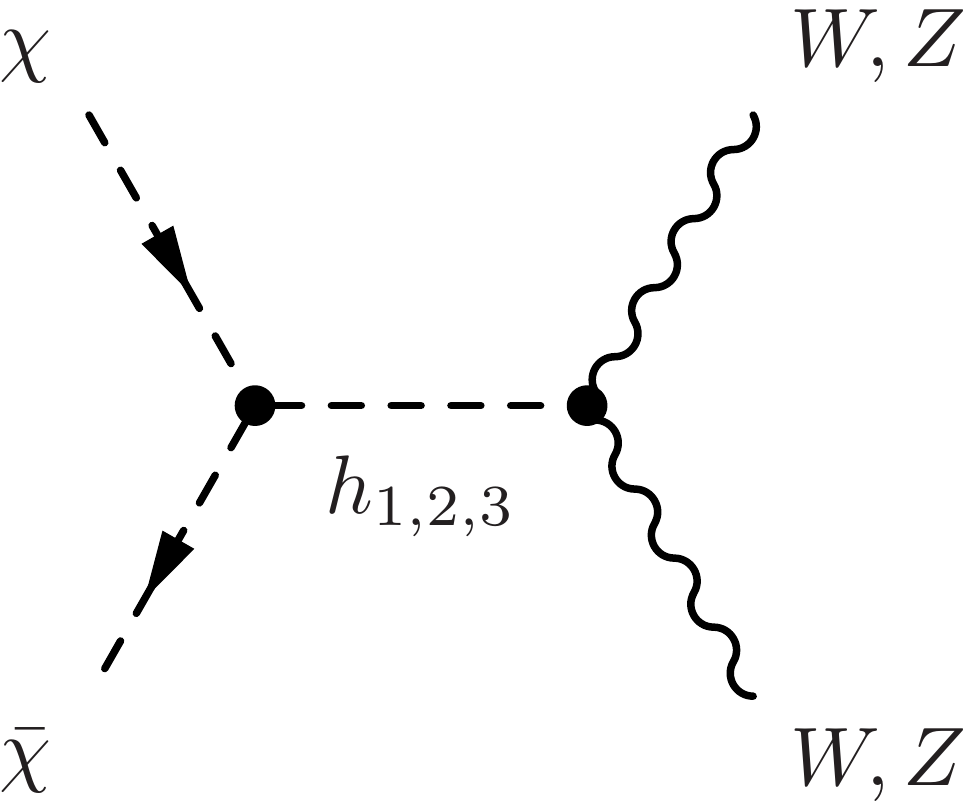}\qquad
  \includegraphics[width=0.21\hsize]{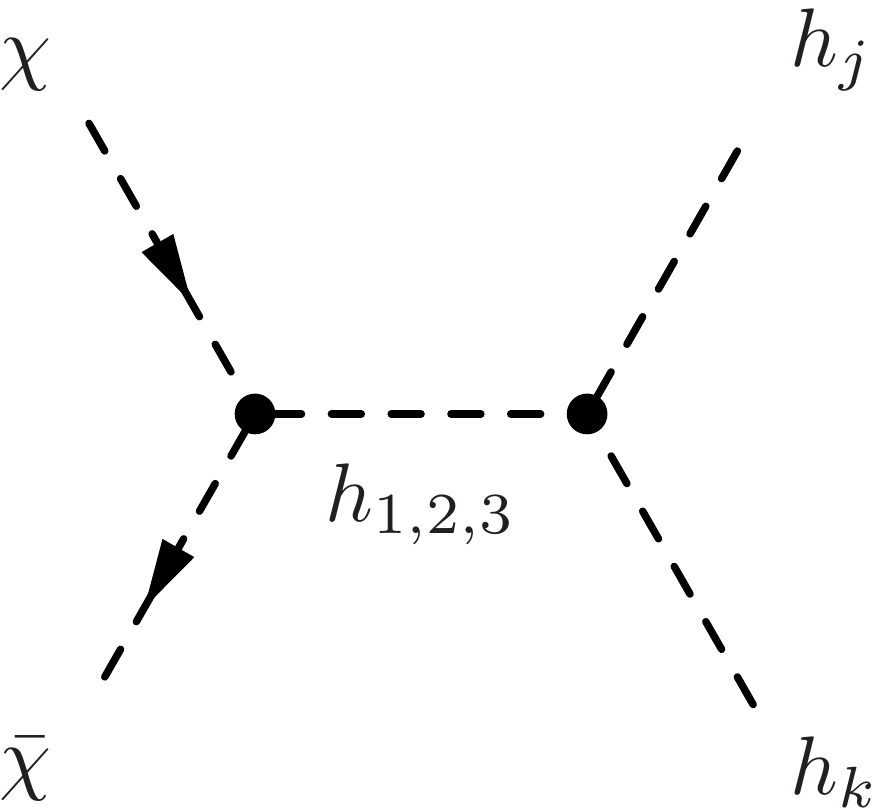} \\ \ \\
  \includegraphics[width=0.23\hsize]{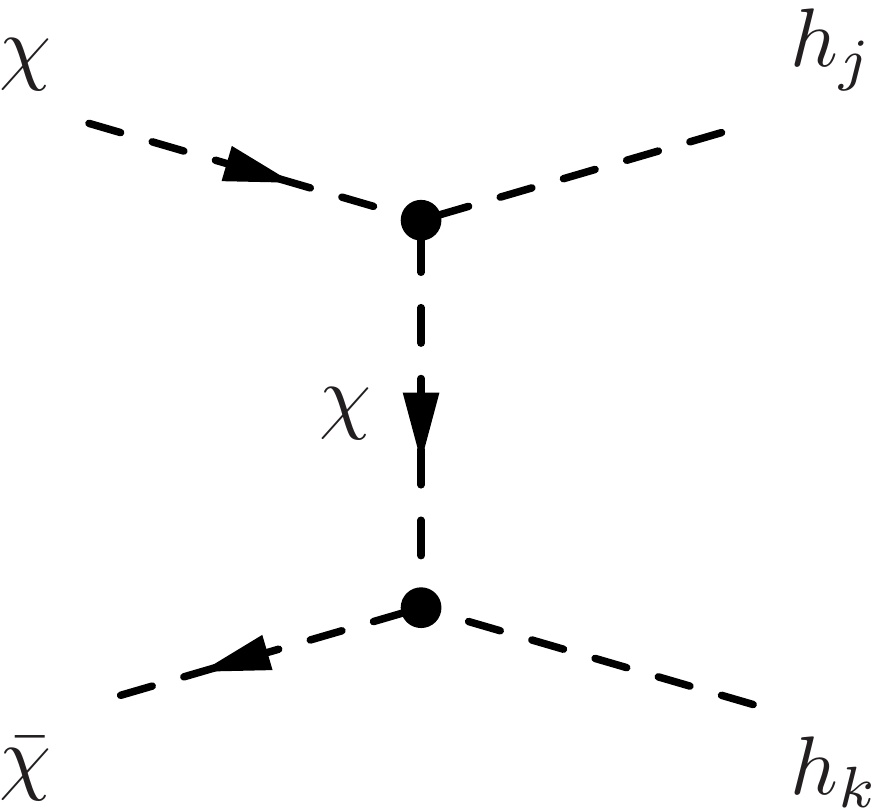}\qquad
  \includegraphics[width=0.23\hsize]{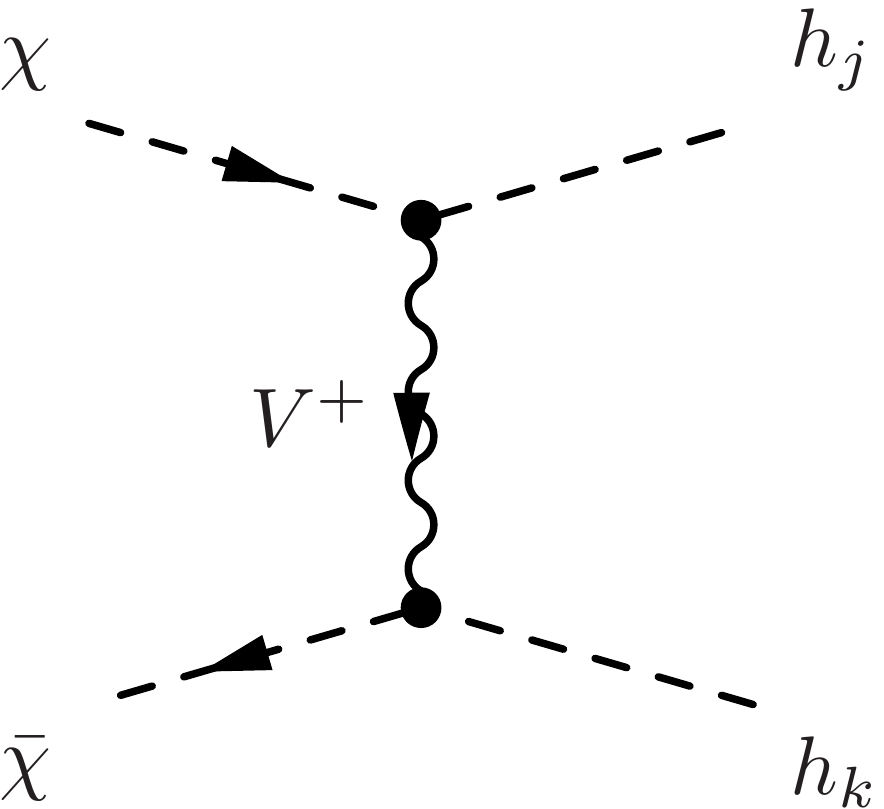} \quad
  \includegraphics[width=0.23\hsize]{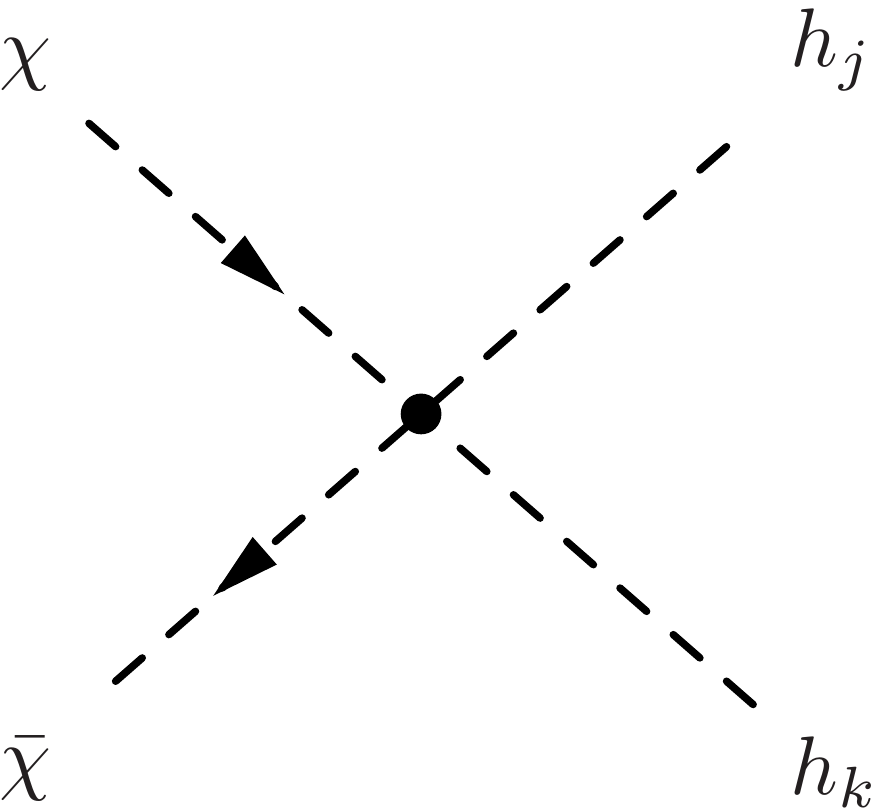}
  \caption{Feynman diagrams for $\chi\bar{\chi}$ annihilation processes in the single-component DM scenario. The arrows on the dashed and wiggly lines show the $U(1)_D$ charge flow.}
  \label{fig:Feynman-diagram-annihilation}
\end{figure}

We discuss the relic abundance of $\chi$ and $\bar{\chi}$.
A pair of $\chi$ and $\bar{\chi}$ annihilates into two SM particles by exchanging $h_j$ in the $s$-channel. 
The processes $\chi \bar{\chi} \to h_j h_k$ are also induced by exchanging $\chi$ or $V^+$ in the $t$- and $u$-channels and by the contact interaction.
Because $m_{V^0} > 2 m_\chi $ and $m_{a_0} > 2 m_\chi$, we do not need to consider processes with $V^{0}$, $V^\pm$, and $a_0$ in the final states.
The relevant Feynman diagrams are shown in figure~\ref{fig:Feynman-diagram-annihilation}.
We calculate the relic abundance by using \texttt{micrOMEGAs}~\cite{Alguero:2023zol} 
and used \texttt{FeynRules}~\cite{Christensen:2008py,Alloul:2013bka} to generate the model file.

\begin{figure}[tbp]
\centering
\includegraphics[width=0.45\hsize]{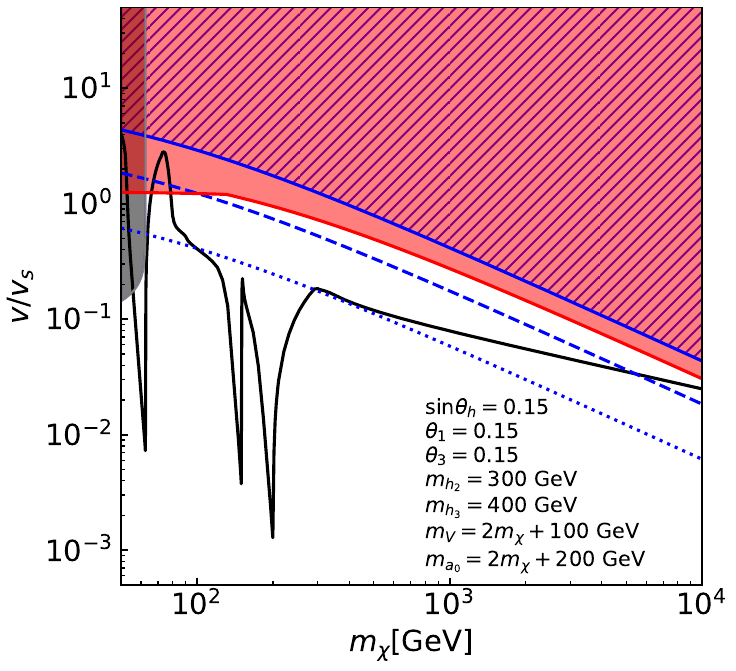}
\includegraphics[width=0.45\hsize]{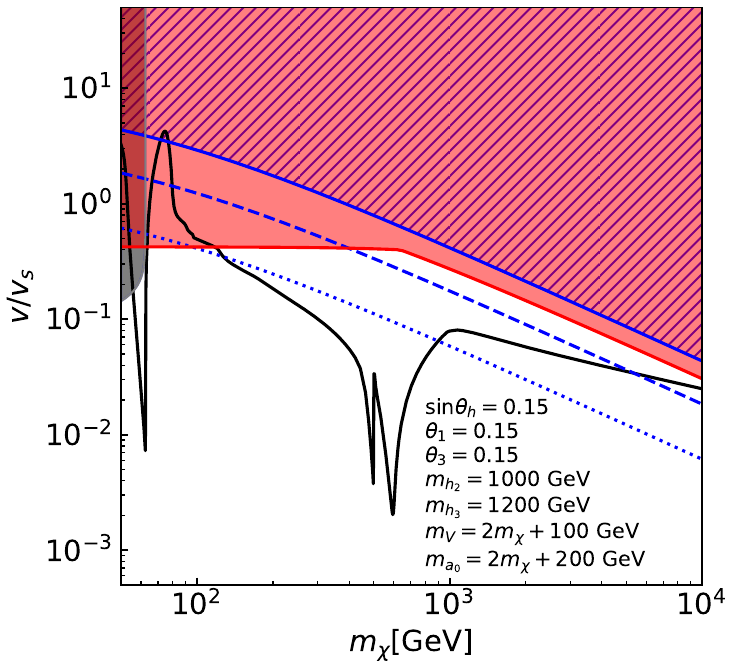}
\caption{Results of the single-component scenario.
The black-solid curves show the value of $v/v_s$ that reproduces the measured value of the DM energy density by the freeze-out mechanism for a given DM mass. Other parameters are shown in each panel.
The black-shaded region for $m_\chi < 62.5$~GeV is excluded by the Higgs invisible decay search.
The red regions are constrained by the PU bound on the scalar couplings.
The PU bound on the gauge coupling constrains the areas above the blue solid curves filled by the hatched pattern. 
On the blue-dashed (blue-dotted) curves, $g_D = 3 \ (1)$.
}
\label{fig:single-component}
\end{figure}

Figure~\ref{fig:single-component} shows the value of $v_s$ that explains the measured value of the DM energy density for a given mass of $\chi$. Here, we take $\sin\theta_h = 0.15$, $\theta_1 = \theta_3 = 0.15$, $m_{V} = 2 m_\chi + 100$~GeV, and $m_{a_0} = 2 m_\chi + 200$~GeV as a benchmark.
In the left (right) panel, we choose $m_{h_2} = 300$~GeV and $m_{h_3} =400$~GeV ($m_{h_2} = 1$~TeV and $m_{h_3} =1.2$~TeV).
We find that the model can reproduce the right amount of the DM relic abundance without contradicting the constraints on the Higgs invisible decay search and the PU bound in most of the regions of the parameter space.
The three dips in each panel are due to the resonance enhancement of $\chi \bar{\chi}$ annihilation by exchanging $h_j$ in the $s$-channel.

In the heavier $m_{\chi}$ regime, the smaller value of $v/v_s$ is required to obtain the right amount of the DM energy density. This $v/v_s$ dependence is different from some other pNG DM models~\cite{Gross:2017dan,2001.03954,Okada:2021qmi,Abe:2022mlc,Otsuka:2022zdy} in which larger values of $v/v_s$ are typically required for heavy DM regime to reproduce the correct relic abundance. This is because all the annihilation processes in other pNG models are mediated by scalar boson exchanging processes or scalar-quartic couplings, and the scalar couplings are required to be sufficiently large for heavy DM to obtain the canonical value of the DM annihilation cross section, $\expval{\sigma v} \simeq 2 \times 10^{-26}$~cm/s. They are given by linear combinations of $m_{h_j}^2/v_s^2$ and 
 $m_{h_j}^2/(v v_s)$. As a result, smaller values of $v_s$, or larger values of $v/v_s$, are necessary. 
On the other hand, in our model, a pair of DM particles can annihilate by exchanging $V^{\pm}$ in the $t$-channel as shown in figure~\ref{fig:Feynman-diagram-annihilation}. The amplitude with $V^\pm$ in the $t$-channel is proportional to the squared of the $SU(2)_x$ gauge coupling, $g_D^2 = (2 m_V/v_s)^2$. Because we take $m_V > 2 m_\chi$ for the single-component DM scenario, it is proportional to $(m_\chi/v_s)^2 = (m_\chi/v)^2 (v/v_s)^2$ for large $m_\chi$, and thus the $V^\pm$ exchanging processes are efficient for heavy DM regime. As a result, larger $v_s$ is required for larger $m_\chi$ to keep the canonical value of $\expval{\sigma v}$. This is the reason why the smaller value of $v/v_s$ is required for heavy DM in our model. 
A similar behavior of $v_s/v$ is also observed in \cite{Abe:2024vxz}.

\subsection{PNG DM scattering off a nucleon}
We discuss the $\chi$-nucleon spin-independent cross section $\sigma_\text{SI}$.
The elementary process is the $\chi$-quark elastic scattering.
At the tree level, it is induced by the exchange of $h_j$ in the $t$-channel. 
Using $g_{\chi \chi h_j}$ and $g_{ffh}$ given in eqs.~\eqref{eq:g_h-chi-chi} and \eqref{eq:g_h-f-f}, 
we find its scattering amplitude as
\begin{align}
 i {\cal M}(\chi q \to \chi q)
=&
-i
\frac{m_q}{vv_s}
\bar{u} u
\sum_{j=1}^3
\frac{m_{h_j}^2 R_{2 j} R_{3j} }{t - m_{h_j}^2} 
\nonumber\\
=&
-i
\frac{m_q}{vv_s}
\bar{u} u
\sum_{j=1}^3
\left(
- R_{2 j} R_{3j} 
+ \frac{t R_{2 j} R_{3j} }{t - m_{h_j}^2} 
\right)
\nonumber\\
=&
-i
\frac{m_q}{vv_s}
\bar{u} u
\sum_{j=1}^3
\frac{t}{t - m_{h_j}^2} R_{2 j} R_{3j}
,
\label{eq:scat-amp-for-xsecSI}
\end{align}
where $u$ and $\bar{u}$ are the Dirac spinors for the quarks.
Here, we use the orthogonality relation, $\sum_{j=1}^3 R_{2j} R_{3j} = 0$.
In the DM direct detection experiments, the momentum transfer is $\mathcal{O}(10)$~MeV, which is much smaller than the masses of the mediators $m_{h_j}$. 
Thus, $t \simeq 0$ is a good approximation for evaluating the spin-independent cross section. 
Because the scattering amplitude given in eq.~\eqref{eq:scat-amp-for-xsecSI} vanishes in $t \to 0$ limit, we conclude that $\sigma_\text{SI} \simeq 0$ at the tree level.\footnote{Here we assume all the CP-even scalars are heavier than $\mathcal{O}(1)$~GeV. If a mediator is lighter than $\abs{t}^{\frac{1}{2}}$, the amplitude of $\chi q \to \chi q$ is not suppressed, and we need a special treatment~\cite{Abe:2021vat}.}

The suppression in eq.~\eqref{eq:scat-amp-for-xsecSI} by the small $t$ is a consequence of a property of NG bosons, soft pion theorem. 
However, $\chi$ is not an exact NG boson but a pNG boson due to the explicit breaking of the global $SU(2)_g$ symmetry by $m_\chi$. 
Hence, the suppression by the small $t$ is not guaranteed at the loop level.
The loop effects on $\sigma_\text{SI}$ in other pNG models are investigated in literature~\cite{Ishiwata:2018sdi, Azevedo:2018exj, Glaus:2020ihj, Biekotter:2022bxp, Abe:2022mlc, Abe:2024vxz}.\footnote{The loop calculation in a model similar to pNG DM models was studied in~\cite{Cho:2023hek}.} 
It was found that the loop induced $\sigma_\text{SI}$ is smaller than the neutrino fog~\cite{Cushman:2013zza} in most of the regions of the parameter space as long as values of relevant couplings satisfy the PU bound. In this paper, we naively expect that the loop induced $\sigma_\text{SI}$ in this model is small. The dedicated study is left for future work.

\section{Two-component DM scenario 1 : \texorpdfstring{$\chi$ and $V^0$}{chi and V0}}\label{sec:chi-V-scenario}
In this section, we investigate a two-component DM scenario where $\chi$ and $V^0$ are the DM candidates. As discussed in section~\ref{sec:DM-candidate}, this scenario is realized for $m_{V^0} < 2 m_\chi$ and $m_{V^0} < m_{a_0}$. Additionally, the stability of $\chi$ requires $m_{\chi} < m_{V^+}$ to forbid the decay process $\chi \to V^+ h_j^{(*)}$. Because $m_{V^0} = m_{V^+}$ at the tree level, it is equivalent to $m_\chi < m_{V^0}$. Therefore, this scenario is realized for $m_\chi < m_V < 2 m_\chi$.

\subsection{Relic abundance}\label{sec:two-component-chi-V}
We calculate the relic abundance of the pNG DM $\Omega_{\chi+\bar{\chi}} h^2$ and the vector DM $\Omega_{V^0}h^2$. There are many annihilation channels. A pair of pNG DM annihilates into not only particles in the thermal bath but also $V^0 h_j^{(*)}$. A pair of $V^0$ can annihilate into $\chi \bar{\chi}$ as well as particles in the thermal bath. Also, co-scattering processes such as $V^0 \chi \to h_j \chi$ are possible. We use \texttt{micrOMEGAs}~\cite{Alguero:2023zol} to calculate the relic abundance. It automatically calculates all the relevant processes for multi-component DM scenarios.

\begin{figure}[tbp]
\centering
\includegraphics[width=0.65\hsize]{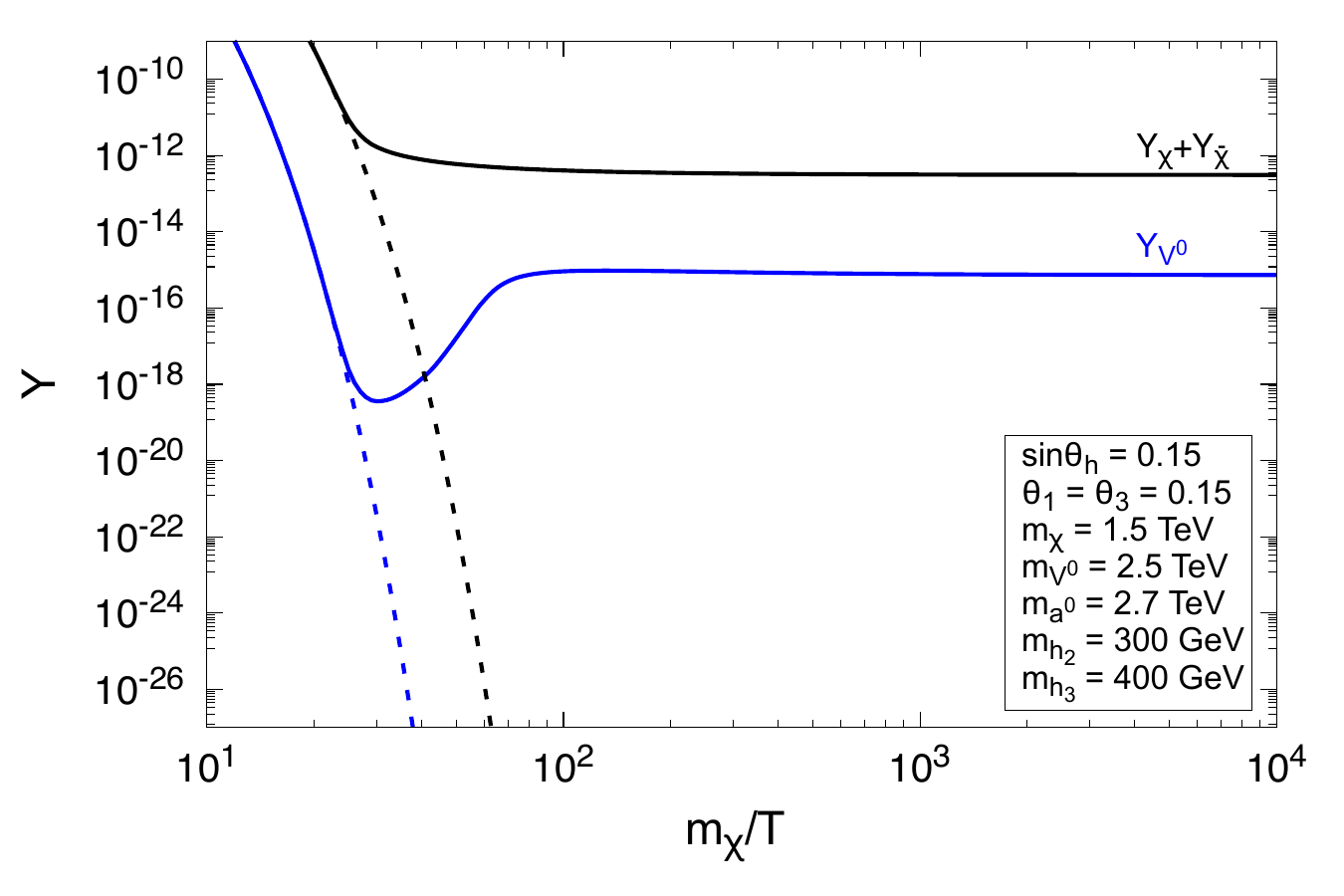}
\caption{Evolution of yields of DM. The black and blue solid curves are for $Y_{\chi} + Y_{\bar{\chi}}$ and $Y_{V^0}$, respectively.
The dashed curves indicate the yields if DM were in the thermal bath. The chosen parameters are also shown.
}
\label{fig:chi-V0_evolution}
\end{figure}

\begin{figure}[tbp]
\centering
\includegraphics[width=0.25\hsize]{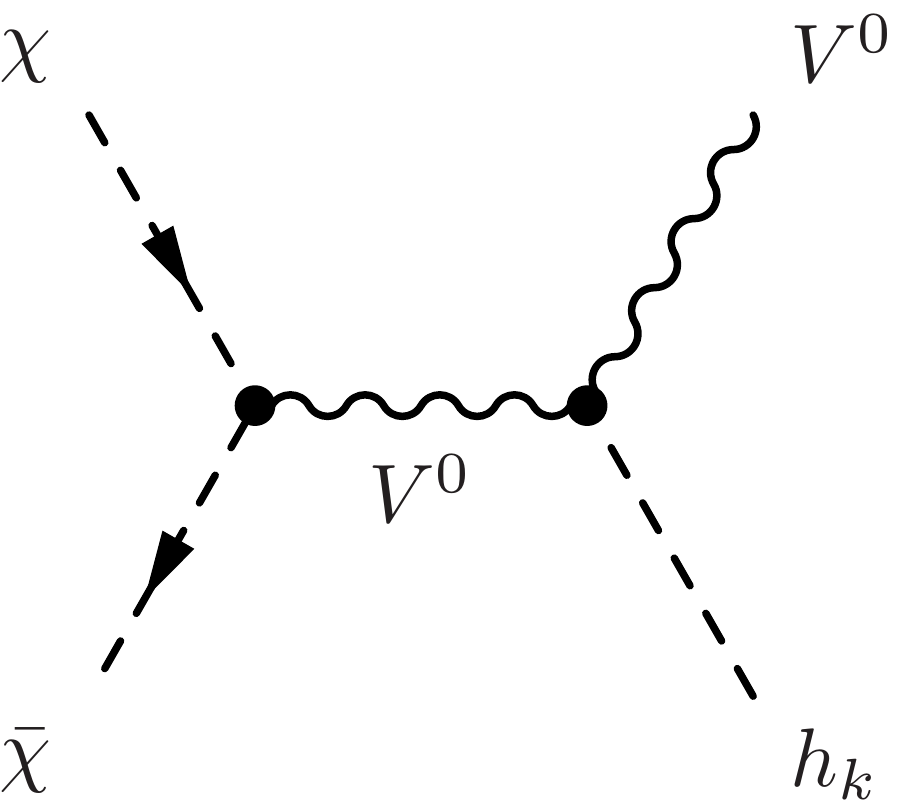} \qquad
\includegraphics[width=0.25\hsize]{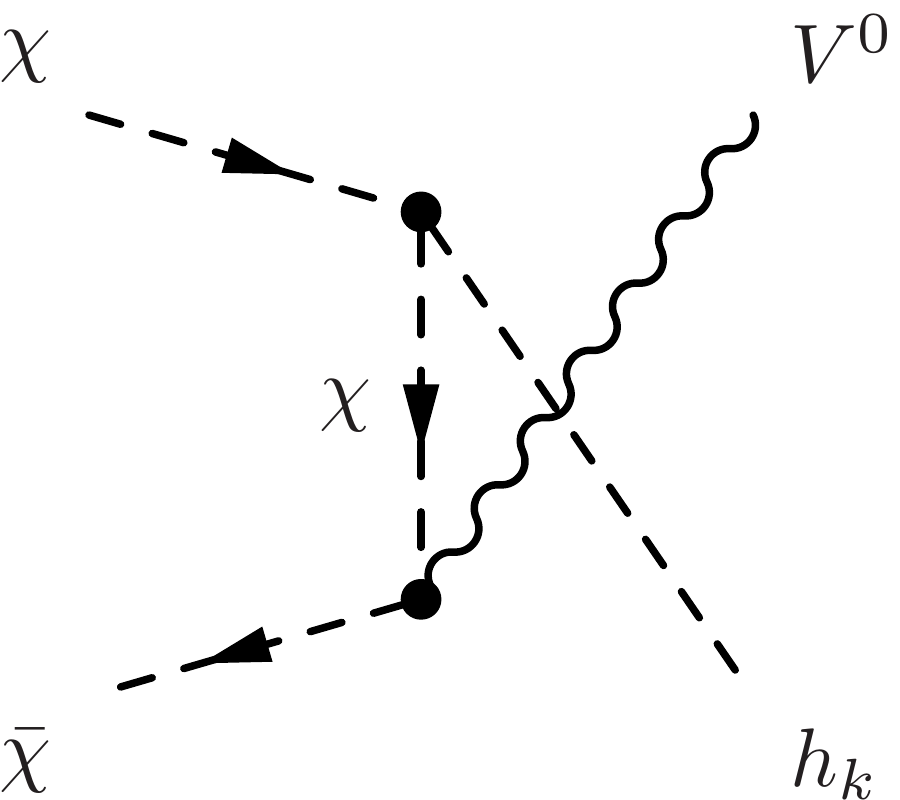} \qquad
\includegraphics[width=0.25\hsize]{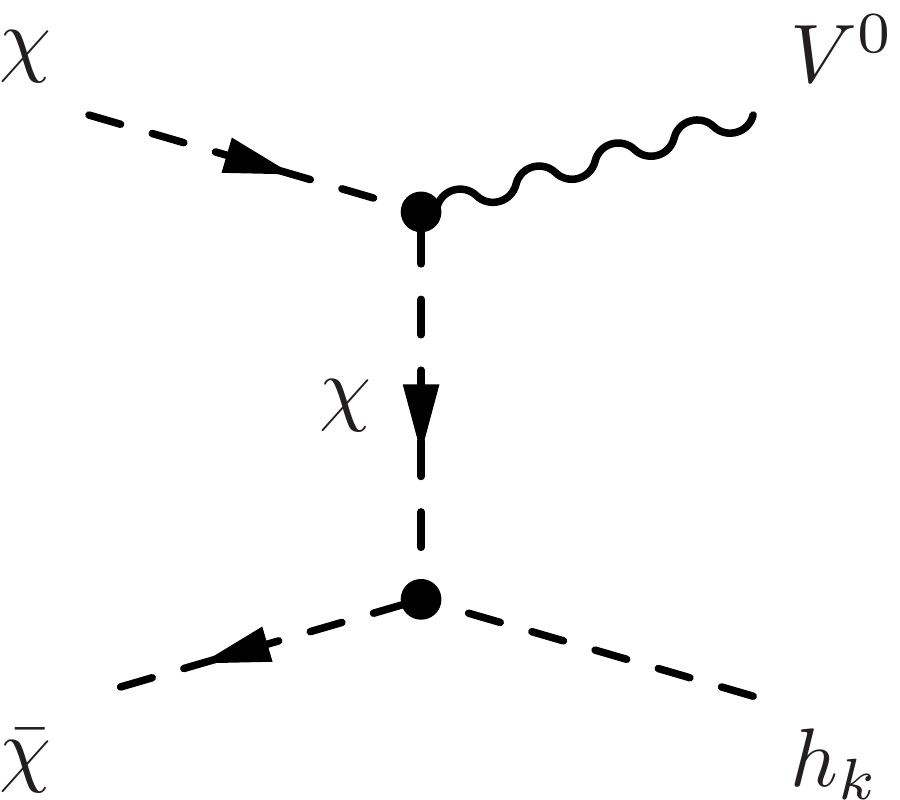}
\caption{Feynman diagrams for $\chi \bar{\chi} \to V^0 h_j.$
}
\label{fig:bouncing-diagrams}
\end{figure}

Figure~\ref{fig:chi-V0_evolution} shows the evolution of the yields $Y$, the ratio of the number density and the entropy density, of $\chi + \bar{\chi}$ and $V^0$.
Here, we take $m_\chi = 1.5$~TeV, $m_{V^0} = $2.5~TeV, $m_{a_0}=2.7$~TeV, $\sin\theta_h = 0.15$, $\theta_1 = \theta_3=0.15$, $m_{h_2}=300$~GeV, and $m_{h_3} =400$~GeV, for the sake of illustration. We choose $v_s = 3.75$~TeV, which reproduces the measured value of the DM energy density, $\Omega_\text{DM} h^2 = \Omega_{\chi + \bar{\chi}}h^2 + \Omega_{V^0}  h^2 = 0.12$.
The vector DM and pNG DM interact with each other via processes such as $\chi \bar{\chi} \to V^0 h_j$ and $\chi V^0 \to \chi h_j$, and they are in equilibrium as long as at least one of them couples to the thermal bath. As a result, the decoupling of the vector and pNG DM from the thermal bath simultaneously happens at $m_\chi/T \simeq 25$.
Just after the decoupling, the ratio of the number densities of $\chi$ and $V^0$ are given by
\begin{align}
 \frac{n_{\chi}}{n_{V^0}}
\sim \frac{n_{\chi}^{eq.}}{n_{V^0}^{eq.}}
\sim \sqrt{\frac{m_\chi^3}{m_{V}^3}} \exp(-(m_\chi - m_V)/T).
\label{eq:neq-pNG_vs_neq-V0}
\end{align}
This implies $n_\chi \gg n_{V^0}$ for $m_\chi < m_V$.
The collision term in the Boltzmann equation consists of the annihilation cross sections multiplied by the number densities of the initial states, such as $\expval{\sigma v}_{\chi V^0\to \chi h_j} n_{\chi} n_{V^0}$. Since $n_\chi \gg n_{V^0}$, the terms proportional to $n_{V^0}$ or $n_{V^0}^2$ are negligible. As a result, the number density of the pNG DM is determined by $\chi \bar{\chi} \to \mathcal{B} \mathcal{B}$, and the number density of the vector DM is determined by $\chi \bar{\chi} \to V^0 h_j$ shown in figure~\ref{fig:bouncing-diagrams}. Here, $\mathcal{B}$ stands for the particles in the thermal bath that consists of
the SM particles and $h_j$. Because of $\chi \bar{\chi} \to V^0 h_j$, $V^0$ is created by a pair annihilation of pNG DM particles, and thus $Y_{V^0}$ increases after the decoupling from the thermal bath as shown in figure~\ref{fig:chi-V0_evolution} for $30 \lesssim m_\chi/T \lesssim 100$.
This increasing behavior is referred to as the bouncing effect~\cite{Katz:2020ywn,Ho:2022erb,Ho:2022tbw,Puetter:2022ucx,DiazSaez:2023wli}.

\begin{figure}[tbp]
\centering
\includegraphics[width=0.47\hsize]{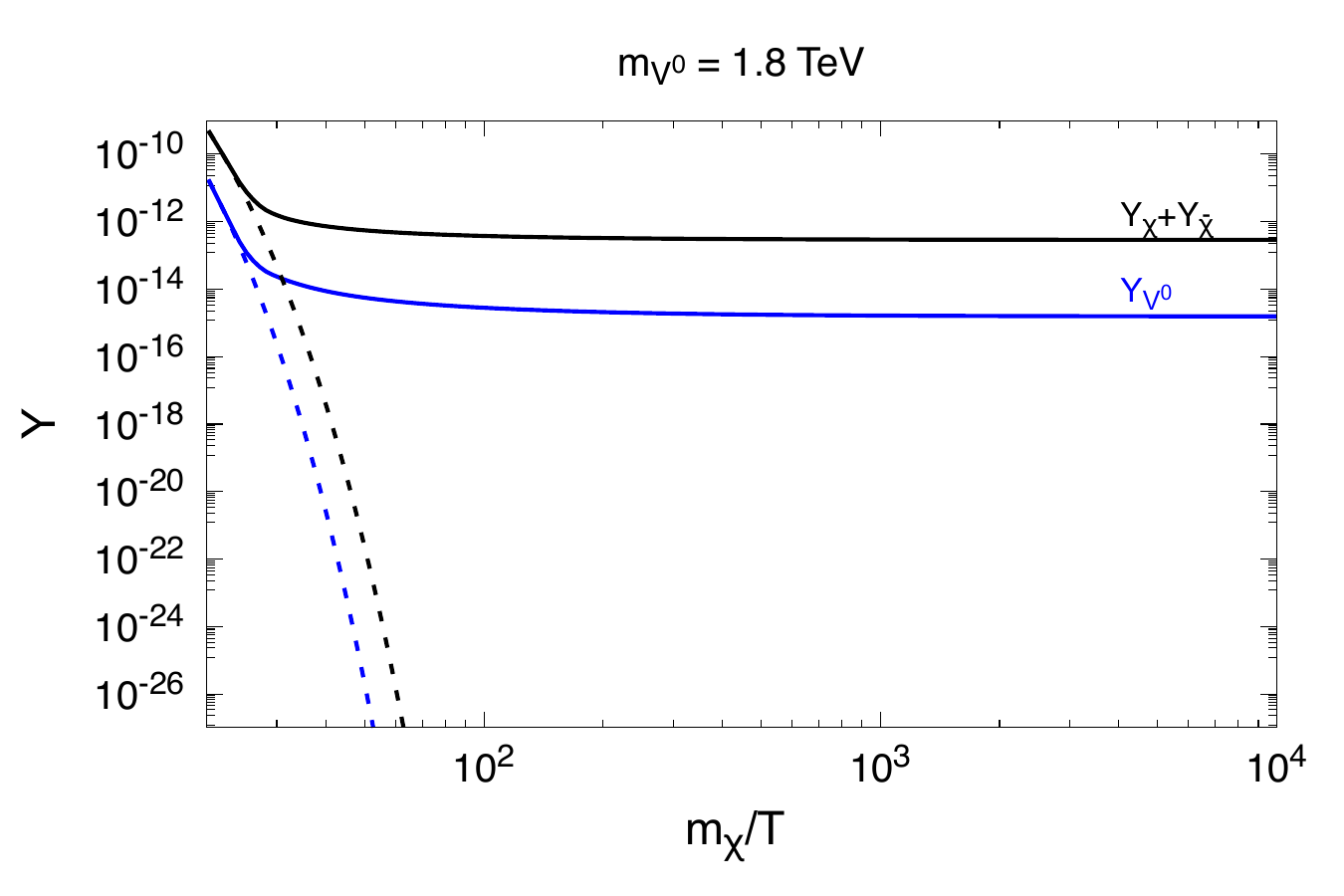}
\includegraphics[width=0.47\hsize]{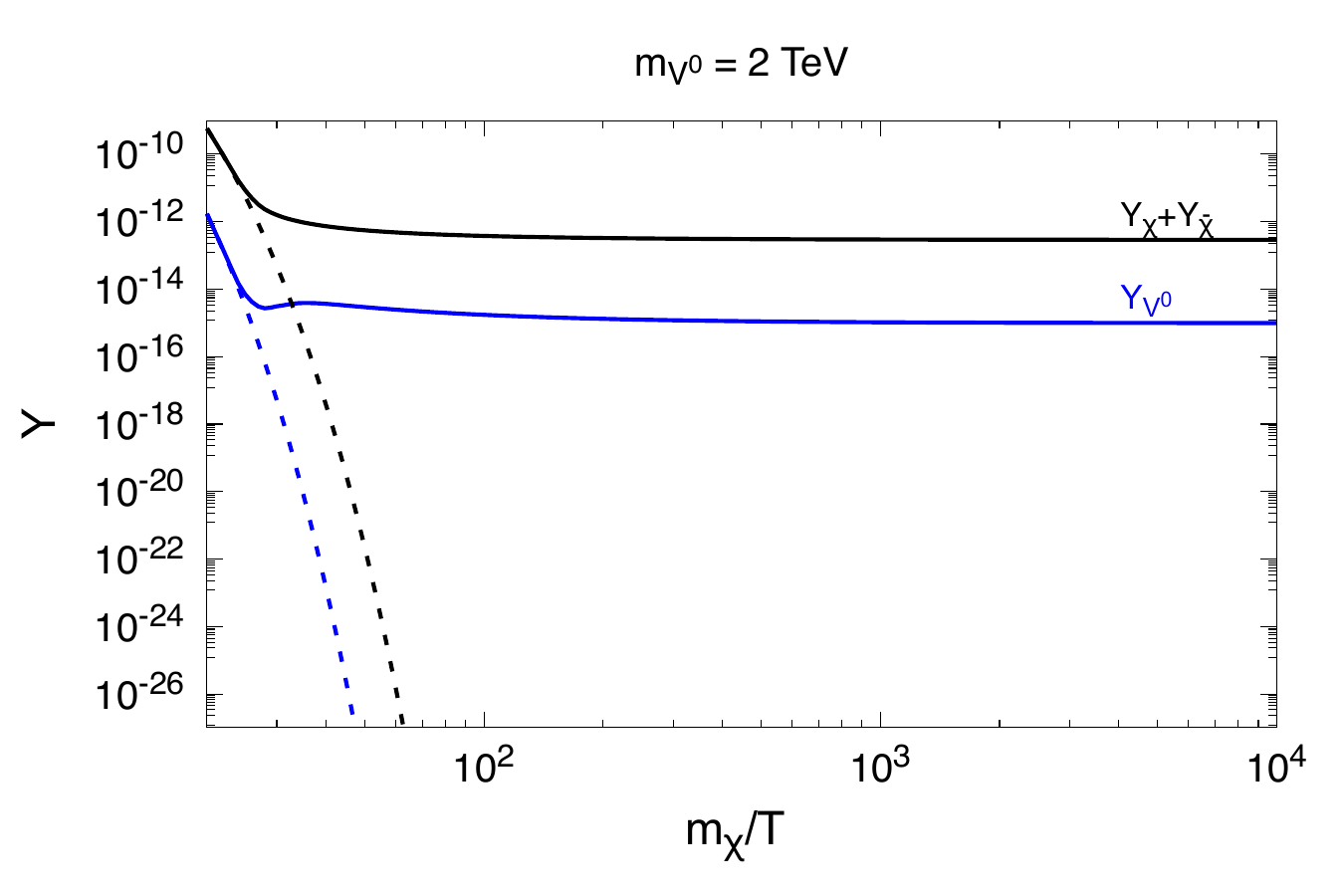}
 \includegraphics[width=0.47\hsize]{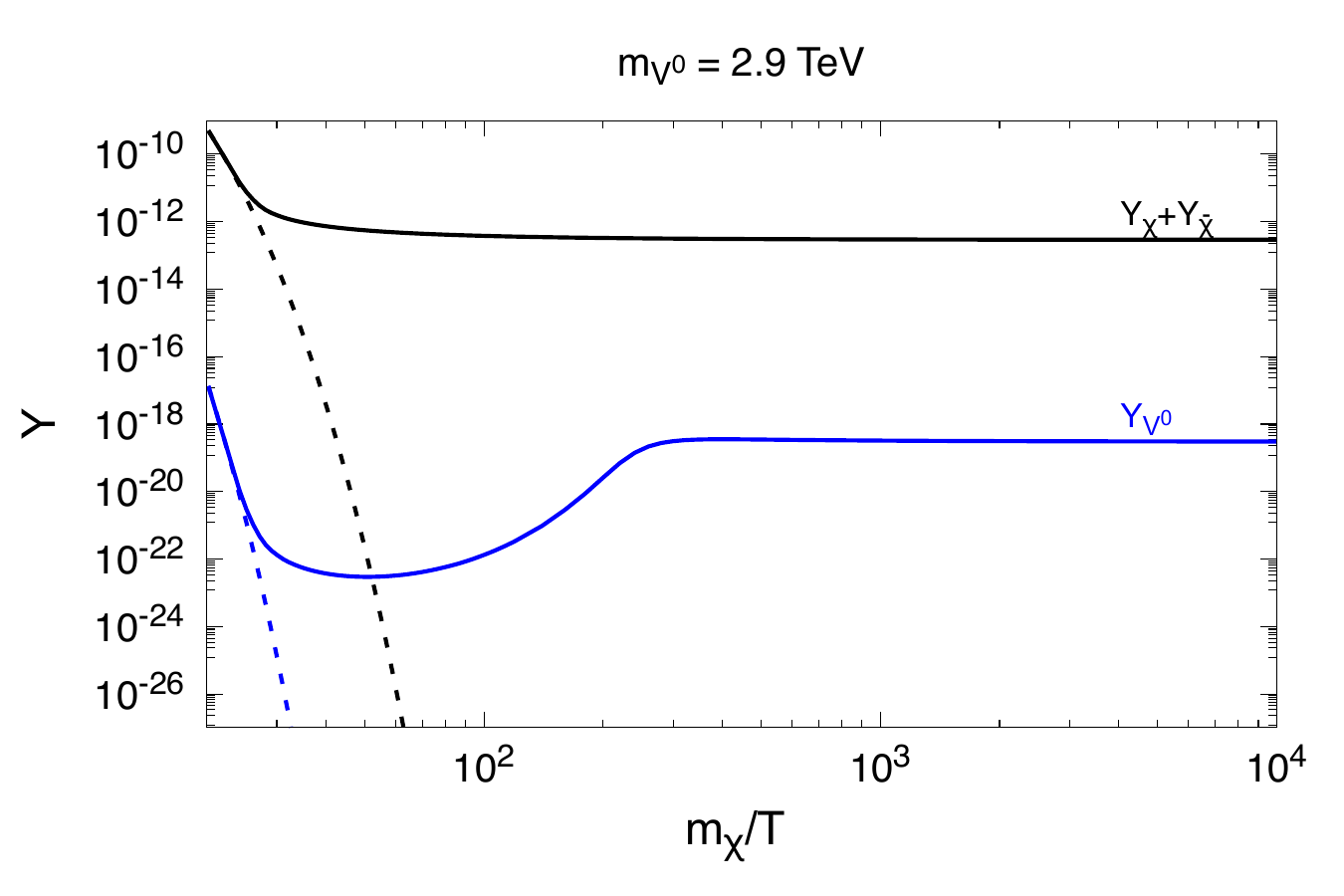}
\includegraphics[width=0.47\hsize]{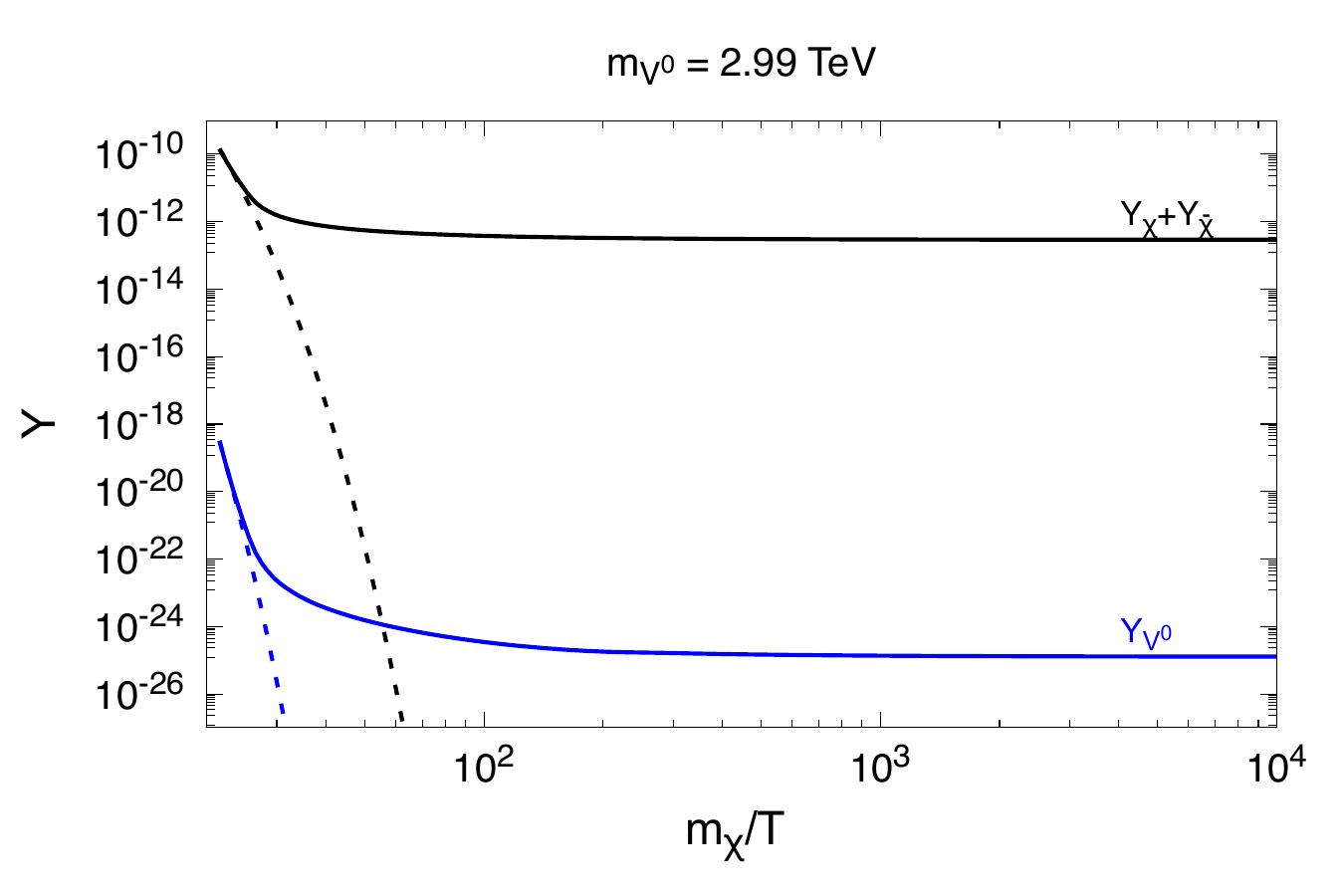}
\caption{
Evolution of $Y$ for $\chi+\bar{\chi}$ (black) and $V^0$ (blue).
The dashed curves are for $Y^{eq.}$.
Here we take $m_\chi = 1.5$~TeV, $m_{a_0} = m_{V} + 200\text{ GeV}$, $\sin\theta_h = 0.15$, $\theta_1 = \theta_3 = 0.15$, $m_{h_2} = 300$~GeV, and $m_{h_3} = 400$~GeV. The values of $m_{V^0}$ are denoted at the top of each panel.
}
\label{fig:x-vs-Y}
\end{figure}

We discuss the $m_{V^0}$ dependence of the bouncing effect.
Figure~\ref{fig:x-vs-Y} is similar to figure~\ref{fig:chi-V0_evolution} but for different values of $m_{V^0}$.
We find that the bouncing effect is significant if the mass difference $m_{V^0} - m_\chi$ is large and the process $\chi \bar{\chi} \to V^0 h_j$ is kinematically allowed.  
As can be seen from the bottom-left panel, the bouncing happens even if $2 m_{\chi} < m_{V} + m_{h_1}$ because $\chi$ and $\bar{\chi}$ obtain kinetic energy from the thermal bath.
However, if $2 m_{\chi} - m_{V} \ll m_{h_j}$, large kinetic energy is necessary for $\chi \bar{\chi} \to V^0 h_j$. It is unlikely to obtain large kinetic energy due to the Boltzmann suppression, and the bouncing behavior disappears, as shown in the bottom-right panel.
For smaller $m_{V^0} - m_\chi$, the ratio of the number densities $n_{V^0}/n_\chi$ is not suppressed, as shown in eq.~\eqref{eq:neq-pNG_vs_neq-V0}. In that case, annihilation processes containing a $V^0$ particle in the initial state, such as $V^0 \chi \to V^+ h_j$, cannot be negligible because their contribution to the collision term in the Boltzmann equation is proportional to $n_{V^0}$. 
In particular, $V^0 \chi \to V^+ h_j$ is $s$-wave but $\chi \bar{\chi} \to V^0 h_j$ is $p$-wave, and thus depletion of $n_{V^0}$ via $V^0 \chi \to V^+ h_j$ is stronger than the creation of $V^0$ via $\chi \bar{\chi} \to V^0 h_j$ for small mass difference regime. As a result, the bouncing behavior disappears in the top panels. 
Note that a $V^+$ particle created via $V^0 \chi \to V^+ h_j$ immediately decays via $V^+ \to \chi h_j$. 
Hence, $V^0 \chi \to V^+ h_j \to \chi h_j h_k$ is regarded as a ``decay'' process of $V^0$, $V^0 \to h_j h_k$, in a medium.

\begin{figure}[tbp]
\centering
\includegraphics[width=0.47\hsize]{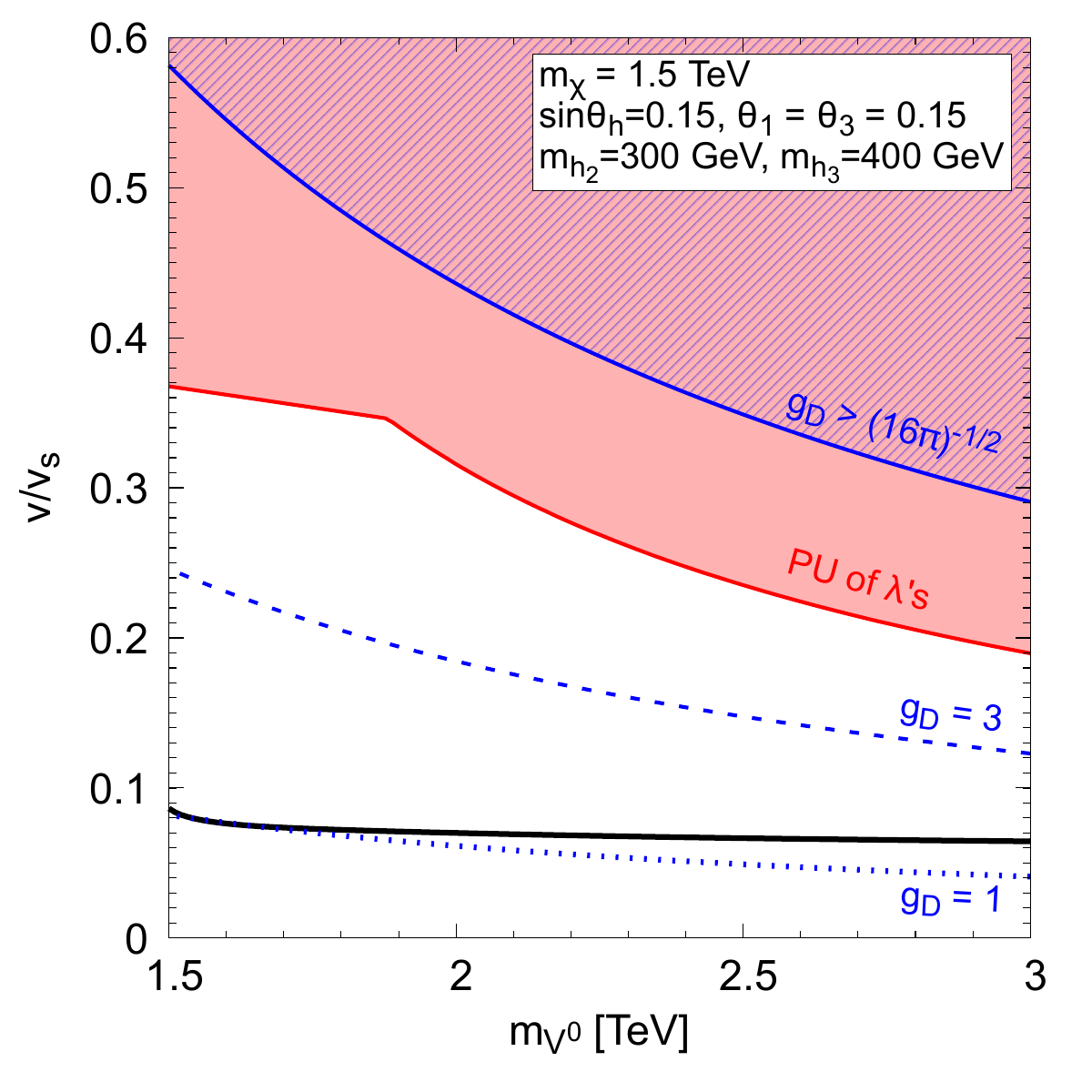}
\caption{
The black-solid curve shows the value of $v/v_s$ that explains the measured value of the DM energy density.
The PU bound is also shown. The red-filled region is excluded by the perturbative unitarity on the scalar quartic couplings.
The blue-hatched region is excluded by the perturbative unitarity on $g_D$. 
The blue-dotted and blue-dashed curves show the contours for $g_D =1$ and $3$, respectively, as a reference.
Here, we take $m_\chi = 1.5$~TeV, $m_{a_0} = m_{V} + 200\text{ GeV}$, $\sin\theta_h = 0.15$, $\theta_1 = \theta_3 = 0.15$, $m_{h_2} = 300$~GeV, and $m_{h_3} = 400$~GeV.
}
\label{fig:chi-V0_vs_mchi1500.pdf}
\end{figure}

Figure \ref{fig:chi-V0_vs_mchi1500.pdf} shows the value of $v/v_s$ that explains the measured value of the DM energy density. 
Here, we vary $m_V$ and set $m_{a_0} = m_{V} + 200\text{ GeV}$. Other parameters are the same as those in figure~\ref{fig:chi-V0_evolution}. The black-solid curve shows the value of $v/v_s$ that reproduces the right amount of the DM energy density. It is almost independent of $m_{V^0}$. This is because the pNG DM is the dominant component of DM, and the relic abundance of the pNG DM is determined by $v_s$ for given $m_\chi$ as discussed in section~\ref{sec:single-component-DM-scenario}. Figure~\ref{fig:chi-V0_vs_mchi1500.pdf} also shows the PU bound. We find that they do not constrain the parameter points for $\Omega h^2 = 0.12$.

\begin{figure}[tbp]
\centering
\includegraphics[width=0.47\hsize]{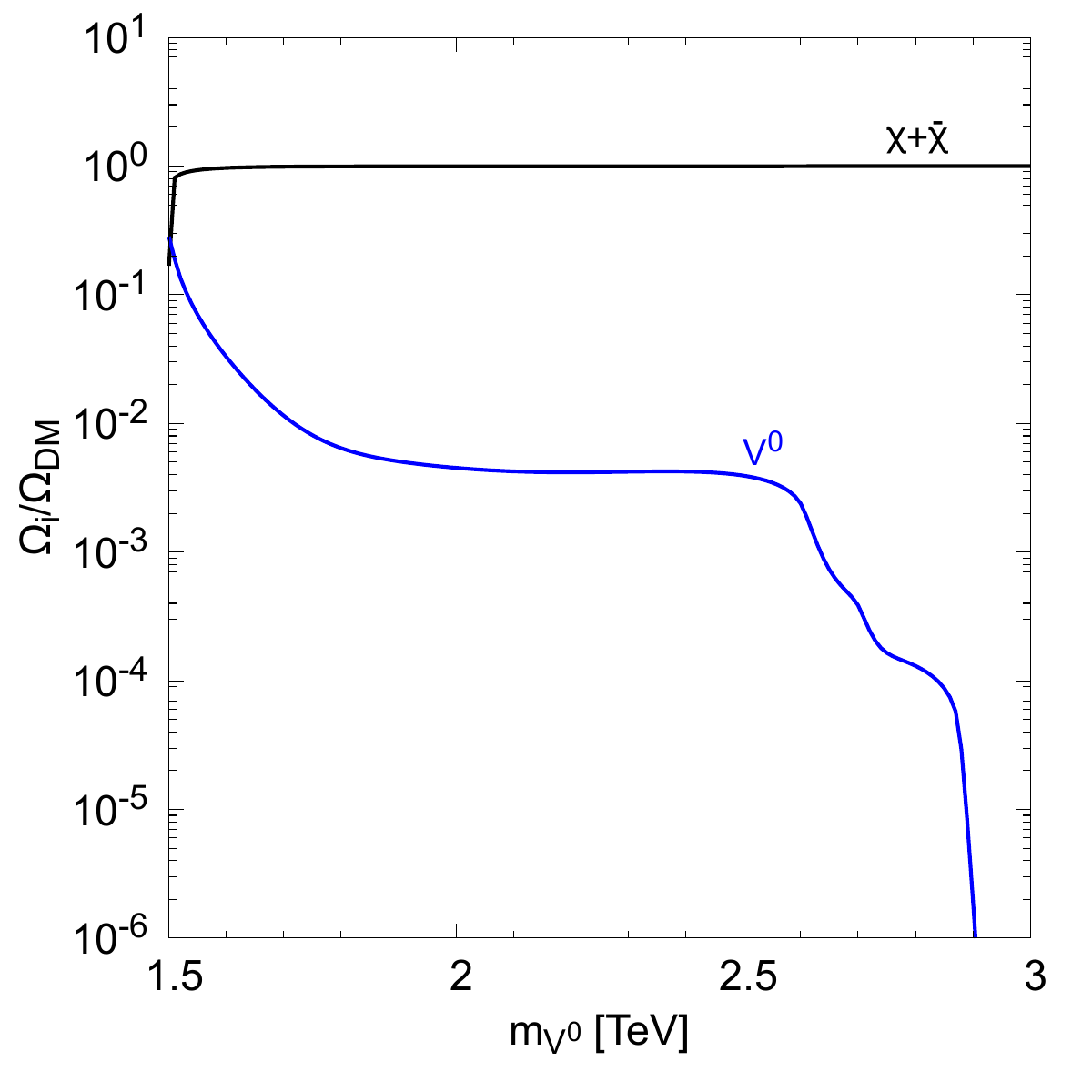}
\includegraphics[width=0.47\hsize]{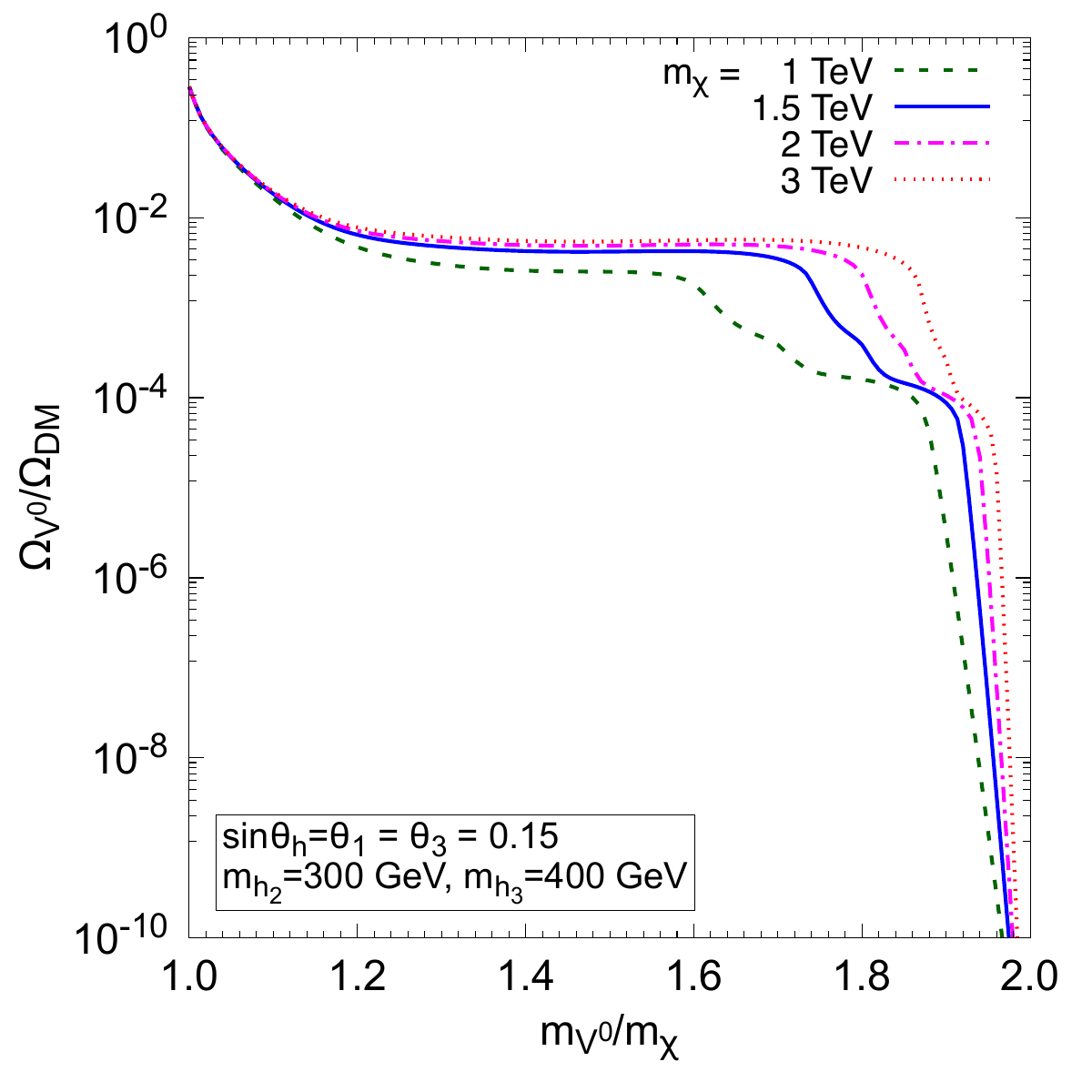}
\caption{
\textbf{Left:} The ratios of each DM energy density to the total DM energy density. 
The black curve is for $\Omega_{\chi + \bar{\chi}} / \Omega_\text{DM}$, and
the blue curve is for $\Omega_{V^0} / \Omega_\text{DM}$, respectively.
Here, we take $m_\chi = 1.5$~TeV, $m_{a_0} = m_{V} + 200\text{ GeV}$, $\sin\theta_h = 0.15$, $\theta_1 = \theta_3 = 0.15$, $m_{h_2} = 300$~GeV, and $m_{h_3} = 400$~GeV.
\textbf{Right:} 
The ratio of the $V^0$ energy density to the total DM energy density as a function of $m_{V^0}/m_\chi$ with different values of $m_\chi$.
The green dashed, blue solid, magenta dotted-dashed, and red dotted curves are for $m_\chi/\text{TeV} = 1$, $1.5$, $2$, and $3$, respectively.
Here we take $m_{a_0} = m_{V} + 200\text{ GeV}$. For other parameters, the chosen values are shown in the bottom-left corner.
}
\label{fig:omegah2-ratio_V0_mchi1500}
\end{figure}

The relic abundance of $V^0$ depends on $m_V$. The left panel of figure~\ref{fig:omegah2-ratio_V0_mchi1500} shows the ratios of each DM energy density to the total DM energy density as a function of $m_{V}$.
Here, the values of the parameters are the same as those in figure~\ref{fig:chi-V0_vs_mchi1500.pdf}. It is clearly shown that the pNG DM is the dominant component of DM. 
In particular, the energy density of $V^0$ drastically changes for $m_{V} > 2 m_\chi - m_{h_3}$. As discussed above, the vector DM is produced via $\chi \bar{\chi} \to V^0 h_j$. It is kinematically allowed for $2 m_\chi > m_{V} + m_{h_j}$. However, for $2 m_\chi < m_{V} + m_{h_j}$, the process is suppressed by the Boltzmann factor because $\chi$ and $\bar{\chi}$ need to obtain kinetic energy from the thermal bath. 
Thus, $\Omega_{V^0}/\Omega_{\text{DM}}$ drastically changes at $m_V = 2 m_\chi - m_{h_3}$, $m_V = 2 m_\chi - m_{h_2}$, and $m_V = 2 m_\chi - m_{h_1}$. 
For $1.2 m_\chi \lesssim m_V \lesssim 2 m_\chi - m_{h_3}$, $\Omega_{V^0}/\Omega_{\text{DM}}$ is almost constant. 
In this mass range, $\Omega_{V^0}$ is determined by $\chi \bar{\chi} \to V^0 h_j$.
For $m_V \lesssim 1.2 m_\chi$, the ratio of the $V^0$ energy density increases as $m_V$ decreases. This behavior also can be seen in figure~\ref{fig:x-vs-Y}. If the mass ratio $m_V/m_\chi$ is close to one, the number density of $V^0$ is not negligible compared to the number density of $\chi$, and thus pair annihilation and semi-annihilation processes of $V^0$ determine the relic abundance of $V^0$. The right panel of figure~\ref{fig:omegah2-ratio_V0_mchi1500} shows $\Omega_{V^0}/\Omega_\text{DM}$ as a function of $m_V/m_\chi$ with different values of $m_\chi$. We find that the behavior changes at $m_{V} \simeq 1.2 m_\chi$, and the the ratio of the $V^0$ energy density to the total DM energy density is almost independent of $m_{V^0}$ for $1.2 m_\chi \lesssim m_V \lesssim 2 m_\chi - m_{h_3}$.

\begin{figure}[tbp]
\centering
\includegraphics[width=0.47\hsize]{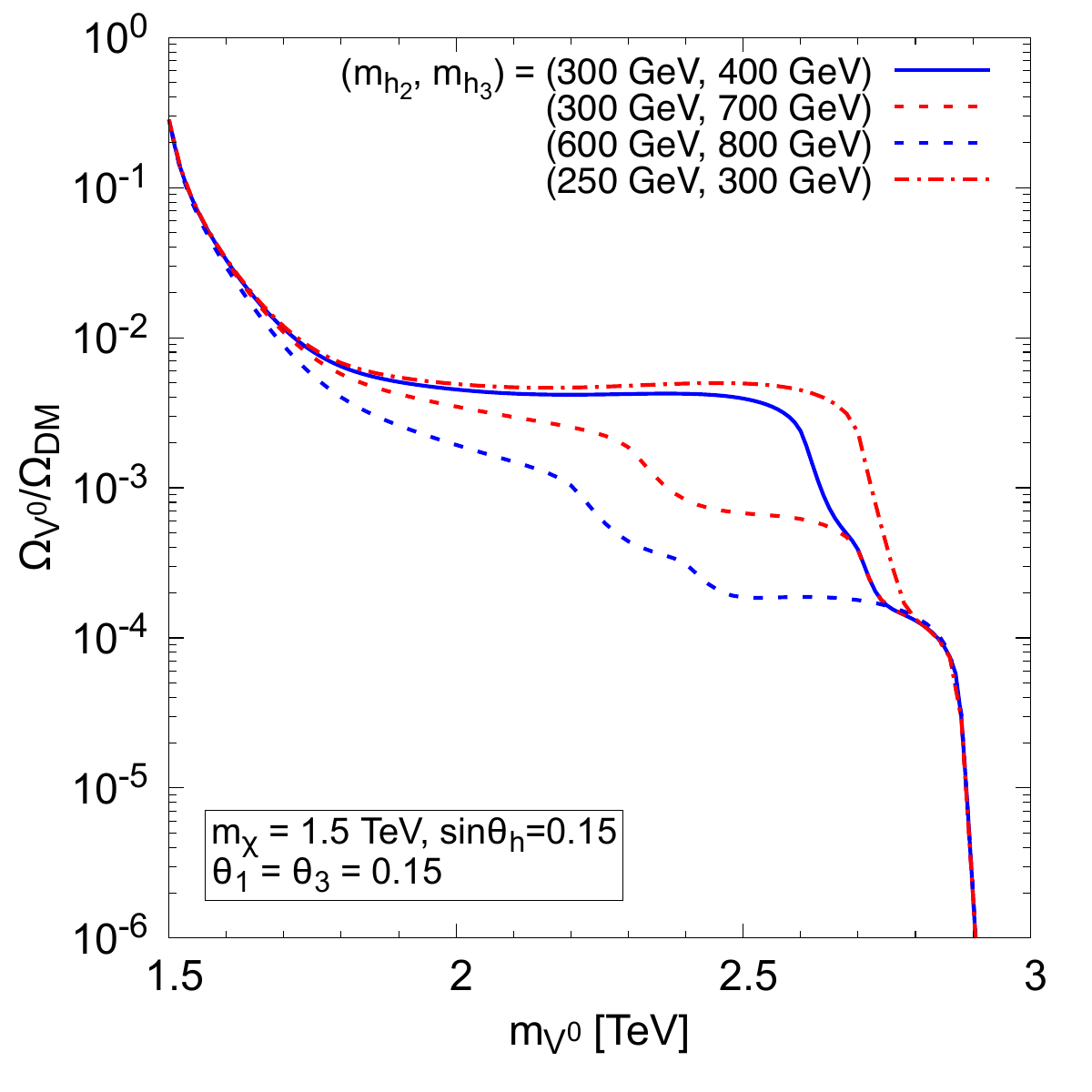}
\includegraphics[width=0.47\hsize]{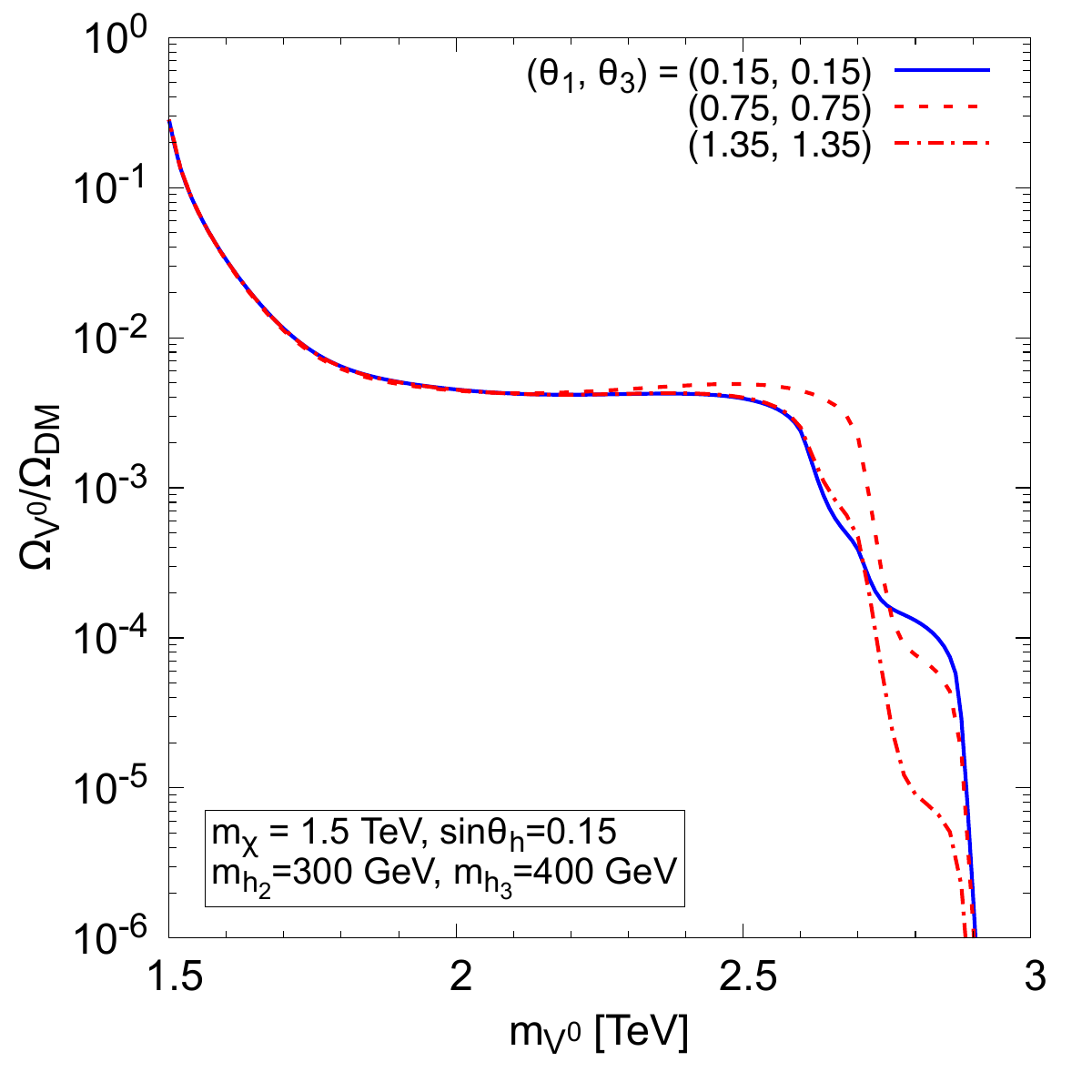}
\caption{
The left (right) panel shows the $m_{h_2}$ and $m_{h_3}$ ($\theta_1$ and $\theta_3$) dependence of the ratio of the $V^0$ energy density to the total DM energy density.
}
\label{fig:V0_comparison}
\end{figure}

The abundance of $V^0$ also depends on $m_{h_j}$ and $\theta_j$. 
We obtain smaller $\expval{\sigma v}_{\chi \bar{\chi} \to V^0 h_j}$ for heavier $h_2$ and $h_3$.  
The left panel of figure~\ref{fig:V0_comparison} shows $\Omega_{V^0}/\Omega_\text{DM}$ for different scalar masses. It clearly shows that the smaller $m_{V^0}$ is required to open the $\chi \bar{\chi} \to V^0 h_2$ ($V^0 h_3$) channel for the heavier $h_2$ ($h_3$). 
In the right panel in figure~\ref{fig:V0_comparison}, the $\theta_{1,3}$ dependence is shown. 
We find that the value of $\theta_{1,3}$ changes the behavior of $\Omega_{V^0}/\Omega_{\text{DM}}$ for $m_{V^0} > 2 m_\chi - m_{h_3}$.
This is because the process $\chi \bar{\chi} \to V^0 h_j$ depends on $\chi\bar{\chi} h_j$ and $V^0 V^0 h_j$ couplings as can be seen from figure~\ref{fig:bouncing-diagrams}, 
and they depend on the scalar mixing angles. 
However, for $m_{V^0} < 2 m_\chi - m_{h_3}$, $\Omega_{V^0}/\Omega_{\text{DM}}$ is almost independent of $\theta_{1}$ and $\theta_3$. The approximated expression of $\expval{\sigma v}_{\chi \bar{\chi} \to V^0 h_j}$ helps the understanding of the behavior. Because $m_{V} > m_\chi \gg m_{h_j}$ in the right panel, $\expval{\sigma v}_{\chi \bar{\chi} \to V^0 h_j}$ for $m_V < 2 m_\chi - m_{h_3}$ is approximately given by 
\begin{align}
 \sum_{j=1}^3 \expval{\sigma v}_{\chi \bar{\chi} \to V^0 h_j}
\simeq&
 \sum_{j=1}^3 \frac{1}{x} \frac{1}{16 \pi m_\chi^2} \frac{m_V^2}{4 m_\chi^2 - m_V^2} \qty(\frac{m_V}{v_s})^4 R_{2j}^2
 \qty( \frac{9}{4} + \frac{m_\chi^2}{m_V^2} ).
\end{align}
Using the orthonormality $\sum_{j=1}^3 R_{2j}^2 = 1$, we find this is independent of $m_{h_j}$ and $\theta_{1,3}$.
Therefore, $\Omega_{V^0}/\Omega_{\text{DM}}$ is insensitive to $\theta_{1}$ and $\theta_3$ for $m_{V^0} < 2 m_\chi - m_{h_3}$.

\subsection{Direct detection}
In contrast to the pNG DM, the scattering amplitude of a vector DM particle off a nucleon is not suppressed by the small momentum transfer. The process is induced by exchanging $h_j$ in $t$-channel. The effective interaction of $V^0$ and a quark $q$ for the $V^0$-nucleon spin-independent scattering cross section $\sigma_\text{SI}^{V^0}$ is given by
\begin{align}
 \mathcal{L}_{V^0q}^{eff.}
=& C_{V^0q} V^{0 \mu} V^0_{\mu} m_q \bar{q}q,
\end{align} 
where
\begin{align}
 C_{V^0q} = - \frac{m_V^2}{v v_s} \sum_j \frac{R_{2j} R_{3j}}{m_{h_j}^2}.
\label{eq:Cvq}
\end{align}
The $V^0$-nucleon spin-independent scattering cross section is given by
\begin{align}
\sigma_\text{SI}^{V^0}
= \frac{1}{\pi} \frac{m_N^4 f_N^2}{(m_V^2 + m_N^2)} C_{V^0 q}^2,
\label{eq:sigmaV}
\end{align}
where $m_N$ is the mass of the nucleon, and 
\begin{align}
 f_N = \frac{2}{9} + \frac{7}{9} \sum_{q=u,d,s} f_q \simeq 0.28.
\end{align}
Here, we use the default values used in \texttt{micrOMEGAs} for $f_q$.

Direct detection experiments give upper bounds on the WIMP-nucleon scattering cross section. 
To obtain the bounds,  they assume that single component DM explains the measured value of the DM energy density. 
To obtain an upper bound on $\sigma_\text{SI}$ in multi-component DM scenarios, we need to rescale the upper bounds obtained by the direct detection experiments according to the number densities of each DM component~\cite{Bhattacharya:2013hva, Abe:2019zhx}.  
In the two-component scenario here, the pNG-nucleon scattering cross section is negligible, and thus the upper bound on $\sigma_\text{SI}^{V^0}$ is given by
\begin{align}
 \frac{\Omega_{V^0}}{\Omega_\text{DM}} \sigma_\text{SI}^{V^0} \equiv \sigma_\text{SI} < \sigma_\text{SI}^{exp.},
\label{eq:sigmaSI-scaling}
\end{align}
where $\sigma_\text{SI}^{exp.}$ is the upper bound on the WIMP-nucleon scattering cross section obtained by direct detection experiments. 
As shown in figure~\ref{fig:omegah2-ratio_V0_mchi1500}, $\Omega_{V^0}/\Omega_\text{DM}$ is less than $\mathcal{O}(1)$\% in a wide region of the parameter space. Consequently, the constraints from the direct detection experiments are weakened, although the $V^0$-nucleon scattering cross section $\sigma_\text{SI}^{V^0}$ itself is not suppressed.

\begin{figure}[tbp]
\centering
\includegraphics[width=0.95\hsize]{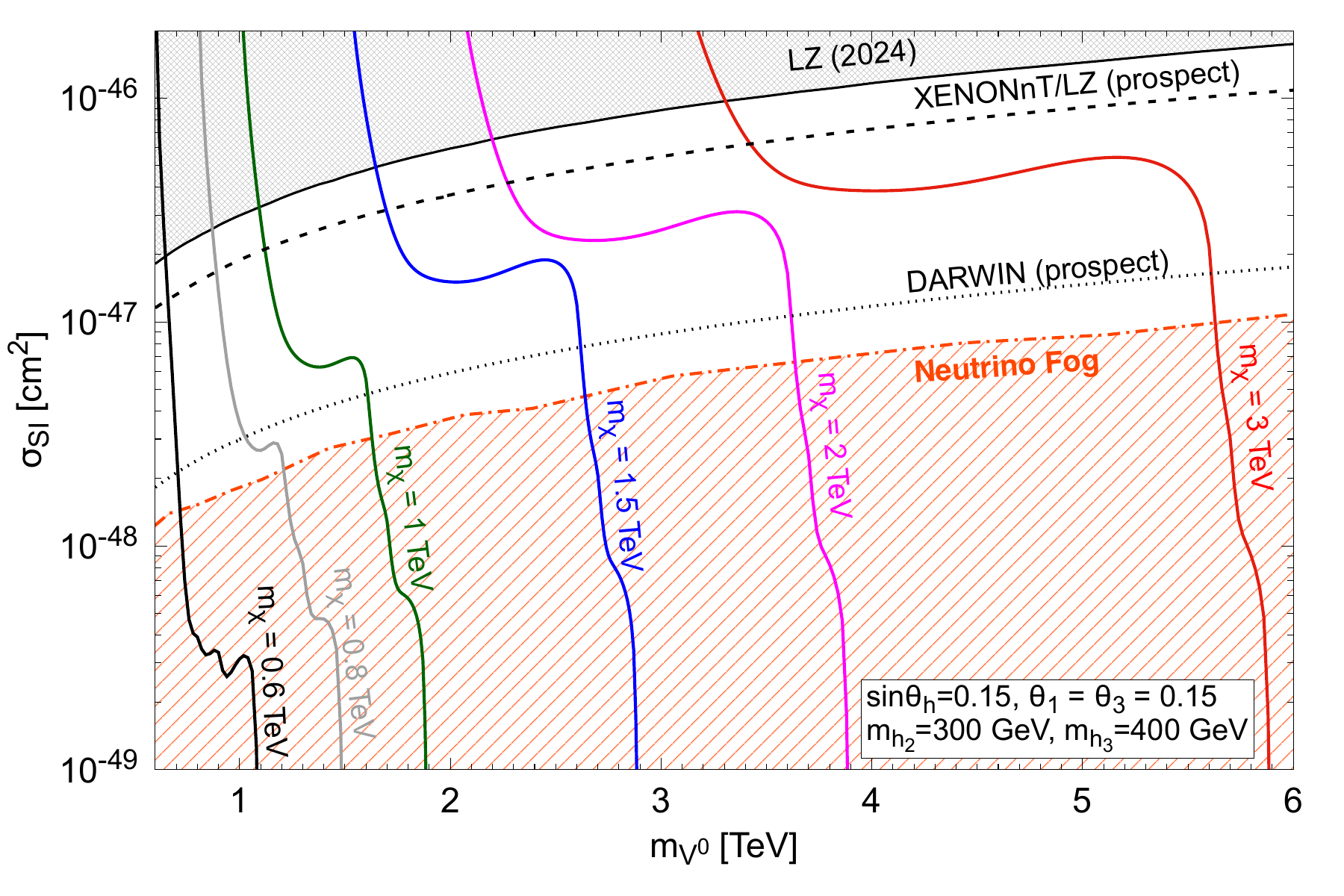}
\caption{
The spin-independent cross section in the $\chi V^0$ scenario. 
The red, orange, blue, green, gray, and black curves show the value of $\sigma_\text{SI}$ as a function of $V^0$ for $m_\chi = 3$~TeV, 2~TeV, 1.5~TeV, 1~TeV, 0.8~TeV, and 0.6~TeV, respectively. 
The cross-hatched pattern shows the region excluded by the LZ experiment~\cite{LZCollaboration:2024lux}.
The dashed and dotted curves are the prospect of the XENONnT/LZ~\cite{XENON:2020kmp, LZ:2018qzl} and DARWIN~\cite{DARWIN:2016hyl} experiments, respectively.
The dot-dashed curve indicates the neutrino fog~\cite{Cushman:2013zza}.
}
\label{fig:chi-V0_sigmaSI_300_400_015}
\end{figure}

Figure~\ref{fig:chi-V0_sigmaSI_300_400_015} shows the value of $\sigma_\text{SI}$ as a function of $m_{V^0}$. 
Here, we take $m_{a_0} = m_{V} + 200\text{ GeV}$, $\sin\theta_h = 0.15$, $\theta_1 = \theta_3 = 0.15$, $m_{h_2} = 300$~GeV, and $m_{h_3} = 400$~GeV as in figure~\ref{fig:omegah2-ratio_V0_mchi1500}. 
We show the cases for $m_\chi = 3$~TeV, 2~TeV, 1.5~TeV, 1~TeV, 0.8~TeV, and 0.6~TeV for the sake of illustration.
Thanks to the suppression by $\Omega_{V^0}/\Omega_\text{DM}$, $\sigma_\text{SI}$ is smaller than the current upper bound by the LZ experiment in most of the region of the parameter space. If the mass of $V^0$ and $\chi$ is degenerate, $\sigma_\text{SI}$ is not suppressed by $\Omega_{V^0}/\Omega_\text{DM}$, and thus the model is already excluded.
We find that $\sigma_\text{SI}$ is larger than the prospect of the DARWIN experiment~\cite{DARWIN:2016hyl} for $m_{V^0} + m_{h_3} < 2 m_\chi$. 
For $m_{V^0} + m_{h_j} > 2 m_\chi$, the relic abundance of $V^0$ significantly small, and $\sigma_{\text{SI}}$ is also highly suppressed and below the neutrino fog. Comparing with figure~\ref{fig:omegah2-ratio_V0_mchi1500}, we find that the behavior of $\sigma_\text{SI}$ is quite similar to $\Omega_{V^0}/\Omega_\text{DM}$ because of eq.~\eqref{eq:sigmaSI-scaling}. For $1.2 m_\chi \lesssim m_V \lesssim 2 m_\chi - m_{h_3}$, $\Omega_{V^0}/\Omega_\text{DM}$ is almost flat but $\sigma_{\text{SI}}$ is tilted. This is because $\sigma_{\text{SI}}$ is proportional to $m_V^2$ as we can see from eqs.~\eqref{eq:Cvq} and \eqref{eq:sigmaV}. Due to the same reason, $\sigma_\text{SI}$ is smaller for smaller $m_{V^0}$. 

As discussed in section~\ref{sec:two-component-chi-V}, $\Omega_{V^0}/\Omega_\text{DM}$ decreases in the heavy $h_2$ and $h_3$ regime, and thus $\sigma_\text{SI}$ is smaller for heavier $h_2$ and $h_3$. 
Also, the $V^0 q$ scattering cross section decreases for large $m_{h_2, h_3}$, as shown in eq.~\eqref{eq:Cvq}. 
Figure~\ref{fig:V0_comparison} also shows the $\theta_j$ dependence of $\Omega_{V^0}/\Omega_\text{DM}$ for $m_V > 2 m_\chi - m_{h_3}$, but $\sigma_\text{SI}$ is smaller in that mass range.
Because larger $\sigma_\text{SI}$ than the neutrino fog is interesting and important to test the model by the forthcoming direct detection experiments, we focus on $m_V < 2 m_\chi - m_{h_3}$,  
where the $\theta_j$ dependence of $\Omega_{V^0}/\Omega_\text{DM}$ is negligible as shown in figure~\ref{fig:V0_comparison}, and do not discuss the $\theta_j$ dependence in $\sigma_\text{SI}$.

We briefly make some comments for the smaller $m_\chi$ case.
In figure~\ref{fig:chi-V0_sigmaSI_300_400_015}, we do not show $m_\chi < 0.6$~TeV 
because the degenerate mass for $m_{V^0}$ and $m_\chi$ is required to obtain larger $\sigma_\text{SI}$ than the neutrino fog. 
Another reason we do not discuss lighter $m_\chi$ is to avoid too large $\sigma_\text{SI}$. 
In the region, $\chi \bar{\chi}$ annihilation processes can hit the $h_{2,3}$ resonance by exchanging $h_{2,3}$ in $s$-channel. It depletes the number density of $\chi$ efficiently, and $V^0$ is the dominant component of DM. Thus, the suppression by the small number density of $V^0$ is not realized, and the direct detection experiments exclude those parameter points.

\section{Two-component DM scenario 2 : $\chi$ and $a_0$}\label{sec:chi-a-scenario}

In this section, we investigate another two-component DM scenario where $\chi$ and $a_0$ are the DM candidates.
As discussed in section~\ref{sec:DM-candidate},
this scenario is realized if $m_{a_0} < m_{V}$, $m_{a_0} < 2 m_\chi$, and $m_\chi < m_{V}$.
In this scenario, $a_0$ is stable due to the $C_\text{dark}$ discrete symmetry, and
$\chi$ is stabilized by the global $U(1)_D$ symmetry.
In contrast to the scenario discussed in section~\ref{sec:chi-V-scenario}, the pNG DM can be heavier than the other DM candidate.

\subsection{Relic abundance}\label{sec:chi-a0_relic}
First, we discuss the energy densities of $\chi+\bar{\chi}$ and $a_0$. 
Similar to the two-component scenario discussed in section~\ref{sec:chi-V-scenario}, there are multiple annihilation channels for the pNG DM and $a_0$.
We use \texttt{micrOMEGAs} to calculate the relic abundance.
The value of $v_s$ is tuned to obtain the right amount of the DM relic abundance, $\Omega_{\chi+\bar{\chi}} h^2 + \Omega_{a_0} h^2 = 0.12$.

\begin{figure}[tbp]
\centering
\includegraphics[width=0.47\hsize]{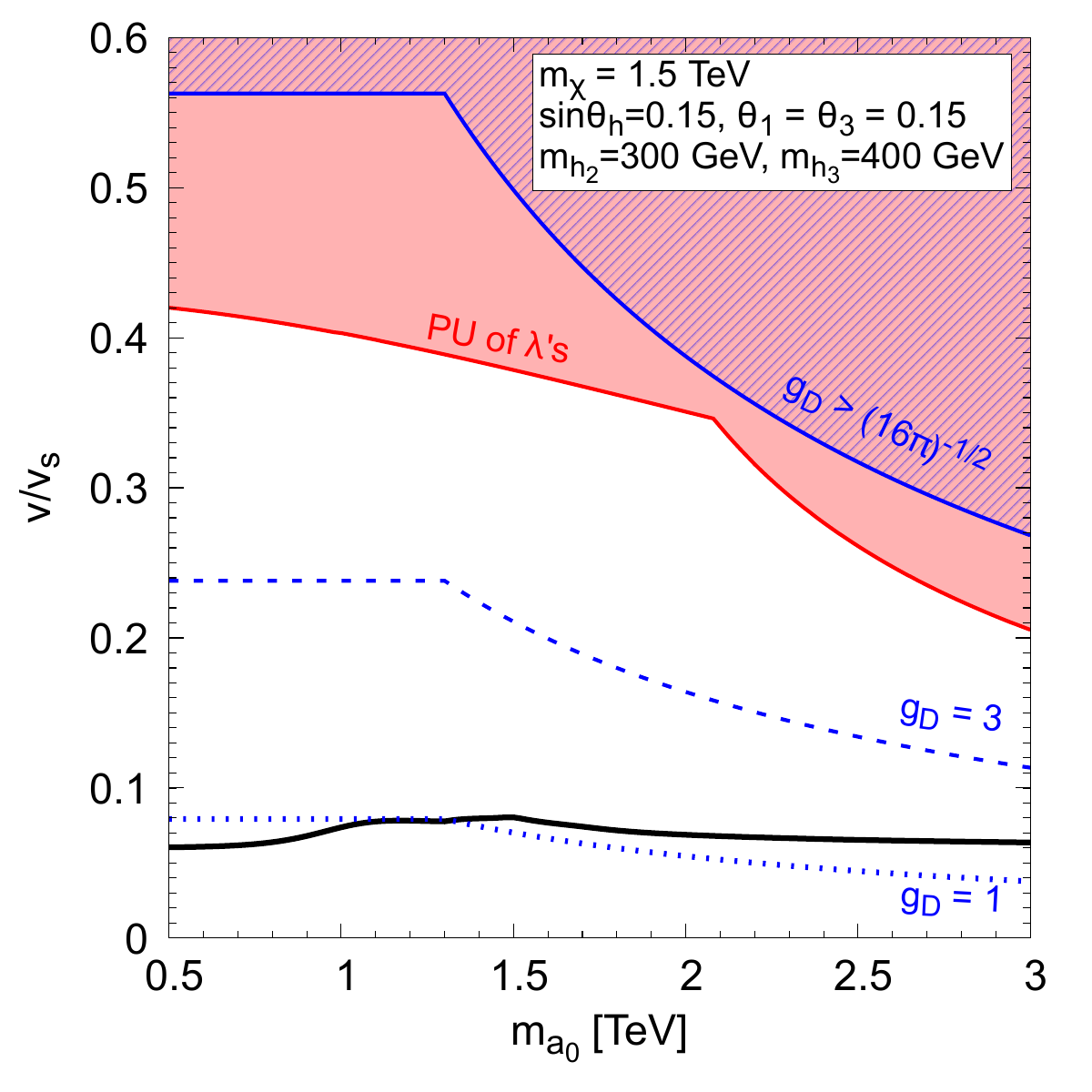}
\includegraphics[width=0.47\hsize]{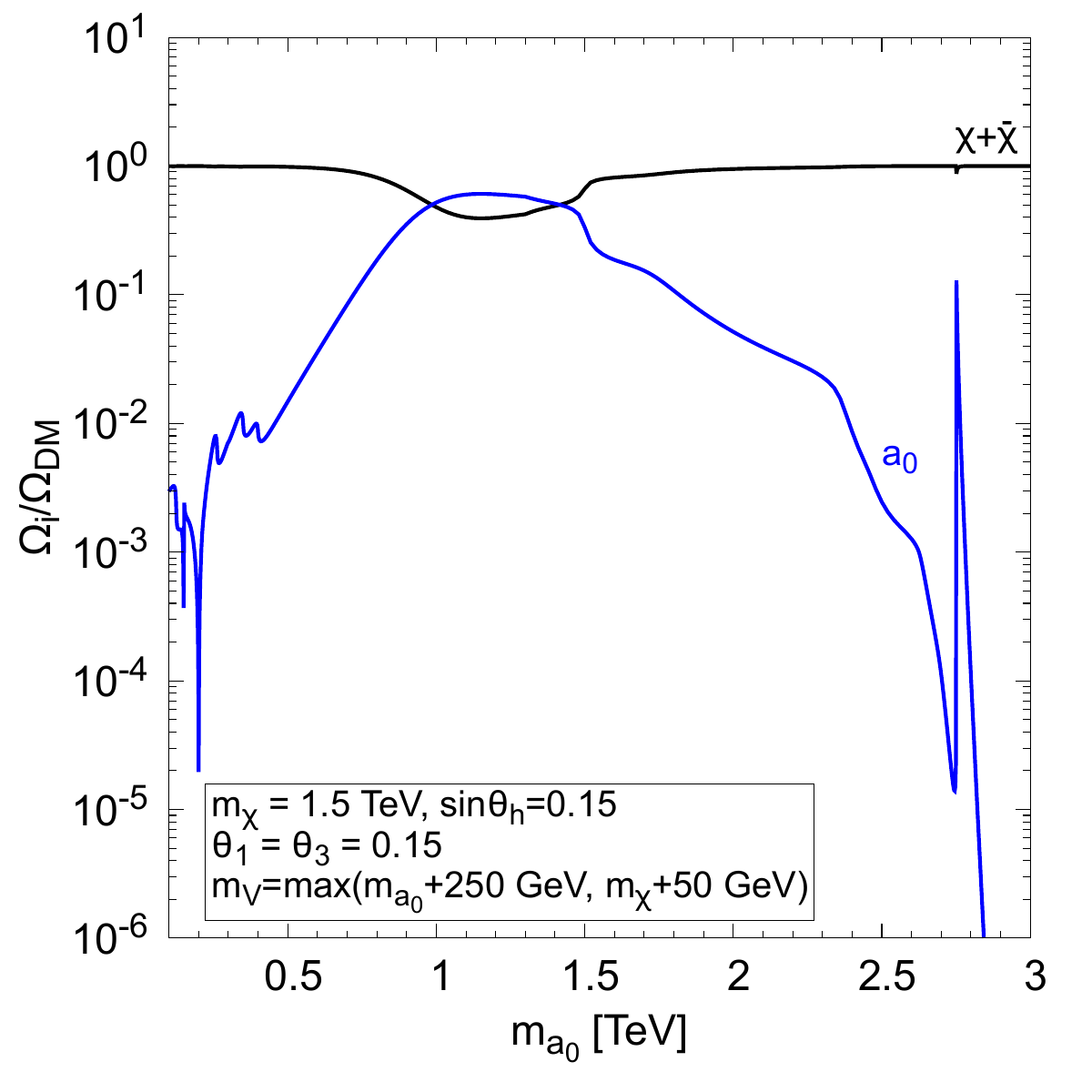}
\caption{
The black-solid curve in the left panel shows the value of $v/v_s$ that reproduces the right amount of the DM relic abundance.
The other color notation is the same as in figure~\ref{fig:chi-V0_vs_mchi1500.pdf}.
Here, we take $m_\chi = 1.5$~TeV, $\sin\theta_h = 0.15$, $\theta_1 = \theta_3 = 0.15$, $m_{h_2} = 300$~GeV, $m_{h_3} = 400$~GeV, and $m_V = \max(m_{a_0} + 250~\text{GeV}, m_\chi + 50~\text{GeV})$.
The right panel shows the ratios of each DM energy density to the total DM energy density. We take the same values for the parameters as in the left panel, and $v_s$ is chosen to obtain $\Omega_\text{DM} h^2 = 0.12$.
The black curve is for $\Omega_{\chi + \bar{\chi}} / \Omega_\text{DM}$, and
the blue curve is for $\Omega_{a_0} / \Omega_\text{DM}$, respectively.
}
\label{fig:vs-and-omegah2-ratio_a0_mchi1500}
\end{figure}
The left panel in figure~\ref{fig:vs-and-omegah2-ratio_a0_mchi1500} shows the value of $v/v_s$ that reproduces the right amount of the DM relic abundance as a function of $m_{a_0}$.
Here, we take $m_\chi = 1.5$~TeV, $\sin\theta_h = 0.15$, $\theta_1 = \theta_3 = 0.15$, $m_{h_2} = 300$~GeV, $m_{h_3} = 400$~GeV, and $m_V = \max(m_{a_0} + 250~\text{GeV}, m_\chi + 50~\text{GeV})$.
The value of $m_V$ is chosen to make $V^{0,\pm}$ unstable.
We find that the value of $v_s$ for $\Omega_\text{DM} h^2 = 0.12$ is almost constant.
The figure also shows that the PU bound does not constrain the parameter points that explain the measured value of the DM energy density.

\begin{figure}[tbp]
\centering
\includegraphics[width=0.47\hsize]{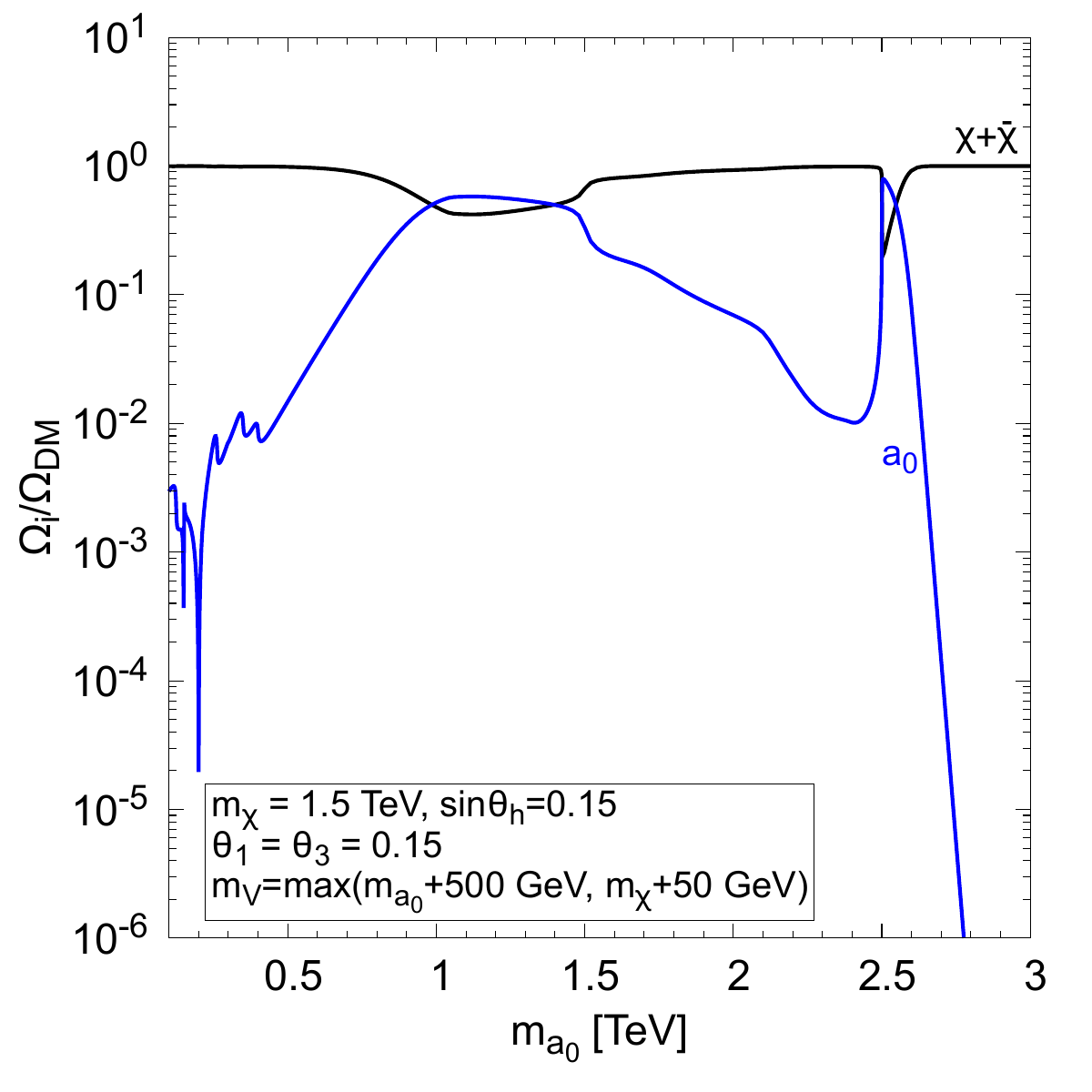}
\includegraphics[width=0.47\hsize]{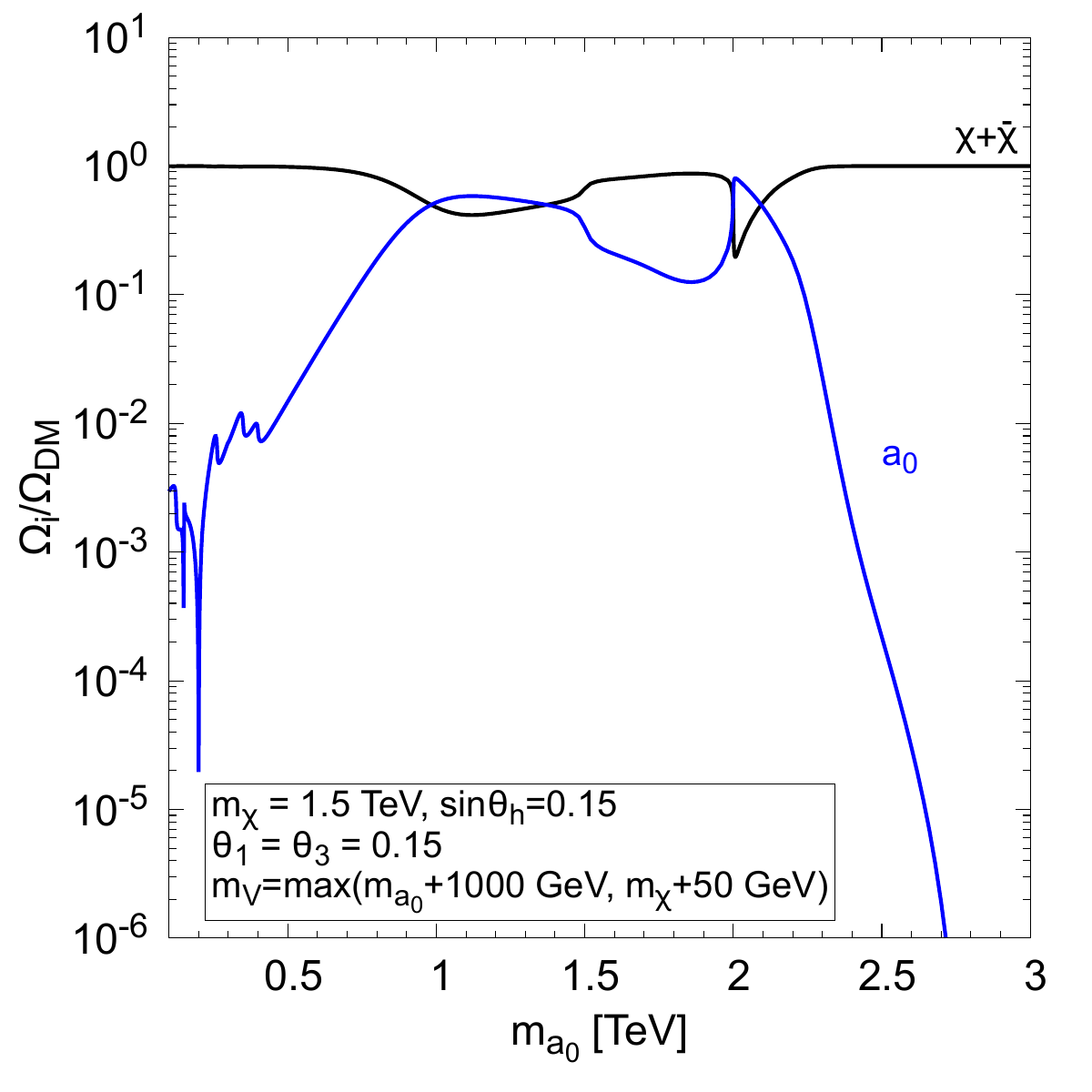}
\caption{
Similar to the right panel of figure~\ref{fig:vs-and-omegah2-ratio_a0_mchi1500} but with different $m_V$, which is shown in each panel. 
}
\label{fig:omegah2-ratio_a0_mchi1500}
\end{figure}

The right panel in figure~\ref{fig:vs-and-omegah2-ratio_a0_mchi1500} shows the ratios of each DM energy density to the total DM energy density,
namely $\Omega_{\chi+\bar{\chi}}/\Omega_\text{DM}$ and $\Omega_{a_0}/\Omega_\text{DM}$, as functions of $m_{a_0}$.  
We take the same values for the parameters as in the left panel, and $v_s$ is chosen to obtain $\Omega_\text{DM} h^2 = 0.12$.
Similar to the $\chi V^0$ scenario discussed in section~\ref{sec:chi-V-scenario}, $a_0$ is mainly produced by the bouncing effect for $m_{a_0} \gtrsim 1.2 m_\chi$.  
The bouncing effect is activated by $\chi \bar{\chi} \to a_0 h_j$ and $\chi \bar{\chi} \to V^0 h_j$. The $V^0$ produced in the later process immediately decays via $V^0 \to a_0 h_j$.
We find a spike at $m_{a_0}= 2.75$~TeV. 
At this point, the mass of $V^0$ is twice the mass of pNG DM, $m_V = 2 m_\chi$. 
As a result, the annihilation process, such as $\chi \bar{\chi} \to V^0 h_j$ and $\chi \bar{\chi} \to a_0 h_j$, are resonantly enhanced by exchanging $V^0$ in the $s$-channel. 
This enhancement efficiently reduces the number density of the pNG DM. Therefore, the number density of $a_0$ increases around this point.
Similar behavior can be seen with different $m_V$ choices. In figure~\ref{fig:omegah2-ratio_a0_mchi1500}, we take $m_V = \max(m_{a_0} + 500~\text{GeV}, m_\chi + 50~\text{GeV})$ in the left panel and
$m_V = \max(m_{a_0} + 1000~\text{GeV}, m_\chi + 50~\text{GeV})$ in the right panel.
We find spikes at $m_{a_0}=$ 2.5~TeV and 2~TeV, where $m_V = 2 m_\chi$, in each panel.

Unlike the $\chi V^0$ scenario, $a_0$ can be lighter than the pNG DM. 
We find that even if the pNG DM is much heavier than $a_0$, it remains the main component, and $a_0$ is the subdominant component.
Since $\chi$ and $a_0$ belong to the same scalar multiplet $\varphi_1$ given in eq.~\eqref{eq:varphi_12}, 
the annihilation cross section of two $a_0$ particles is of the same order as that of $\chi$ and $\bar{\chi}$ in a naive estimate. 
However, they differ because of the explicit breaking of a global $SO(3)$ symmetry by $\lambda_3 \neq \lambda_4$.
We discuss the detail of the global symmetry in section~\ref{sec:DD_chi-a} and appendix~\ref{sec:enhanced-symmetry}.
The breaking generates the mass difference between $\chi$ and $a_0$, and 
their couplings are also different by the mass difference. For example, $g_{\chi \chi h_j} = g_{aah_j} + 2(m_{a_0}^2 - m_\chi^2) R_{2j}/ v_s$, see eqs.~\eqref{eq:g_h-chi-chi} and \eqref{eq:g_h-a-a}.
The large mass difference between $a_0$ and $\chi$ makes the $a_0$ couplings typically larger than the $\chi$ couplings, and thus the annihilation cross section of a pair of $a_0$ particles are also larger than that of the pNG DM. As a result, the relic abundance of $a_0$ is smaller than that of $\chi$ and $\bar{\chi}$ for large mass different regions.

\subsection{Direct detection}\label{sec:DD_chi-a}
Next, we discuss $\sigma_\text{SI}$. 
The elastic scattering of a $a_0$ particle off a nucleon is induced by diagrams exchanging $h_j$ in $t$-channel at the tree level.  
The effective $a_0$-$q$ interaction and the $a_0$-nucleon spin-independent scattering cross section $\sigma_\text{SI}^{a_0}$ are given by
\begin{align}
 \mathcal{L}_{a_0q}^{eff.}
=& C_{a_0 q} a_{0}^2 m_q \bar{q}q, \\
\sigma_\text{SI}^{a_0}
=& \frac{1}{\pi} \frac{m_N^4 f_N^2}{(m_{a_0}^2 + m_N^2)} C_{a_0 q}^2,
\label{eq:sigma_a}
\end{align} 
where
\begin{align}
 C_{a_0 q} = \frac{2 (m_{a_0}^2 - m_\chi^2)}{v v_s} \sum_j \frac{R_{2j} R_{3j}}{m_{h_j}^2}.
\label{eq:Caq}
\end{align}
Similar to the $\chi V^0$ scenario in section~\ref{sec:chi-V-scenario}, we rescale $\sigma_\text{SI}^{a_0}$ by 
\begin{align}
 \sigma_\text{SI}= \frac{\Omega_{a_0}}{\Omega_\text{DM}} \sigma_\text{SI}^{a_0},
\end{align}
and compare it with the direct detection experiments. 
Note that $\sigma_\text{SI}^{a_0}$ in eq.~\eqref{eq:sigma_a} is not suppressed by the small momentum transfer, and thus small $\Omega_{a_0}/\Omega_\text{DM}$ is required to explain the current null results in the direct detection experiments.

\begin{figure}[tbp]
\centering
\includegraphics[width=0.47\hsize]{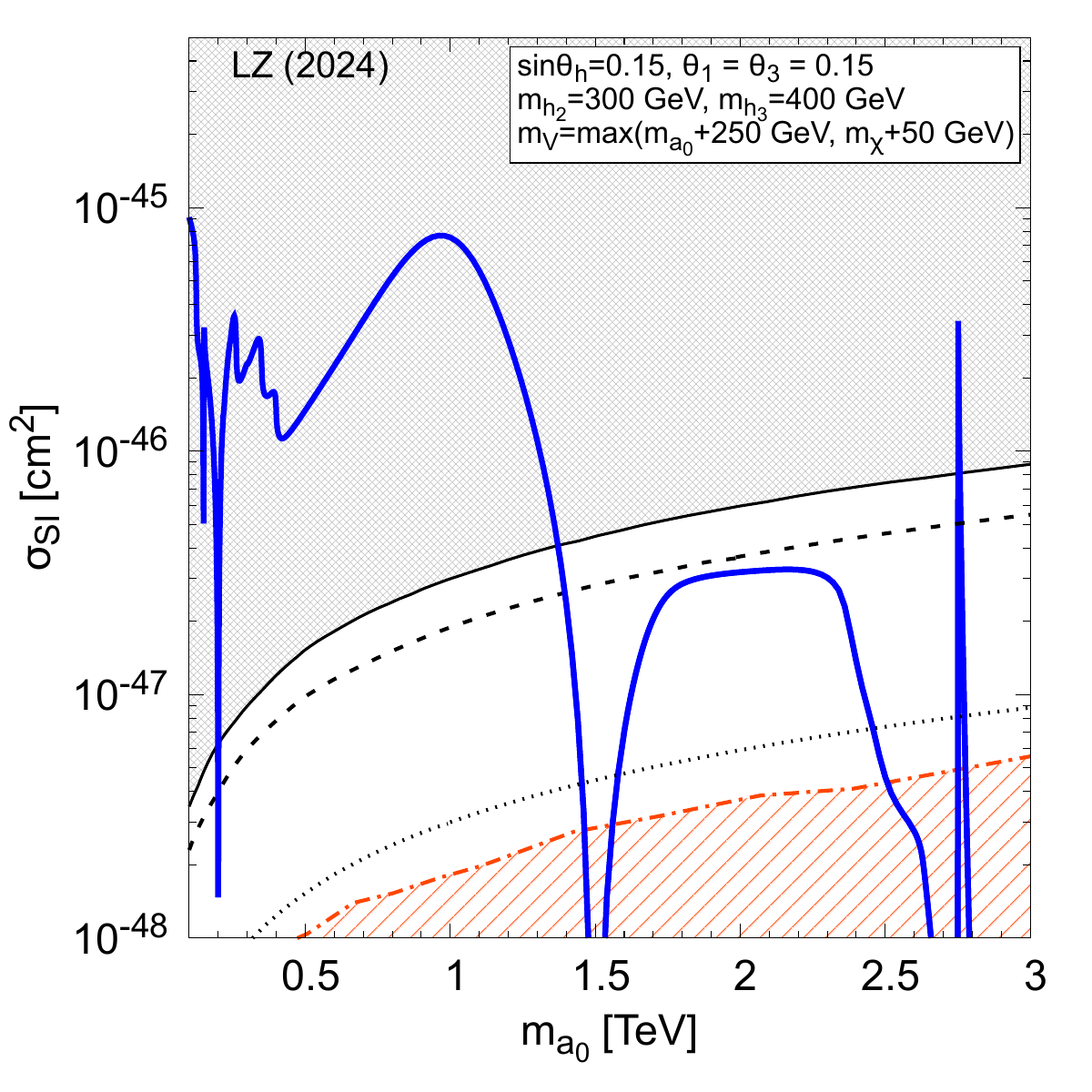}
\includegraphics[width=0.47\hsize]{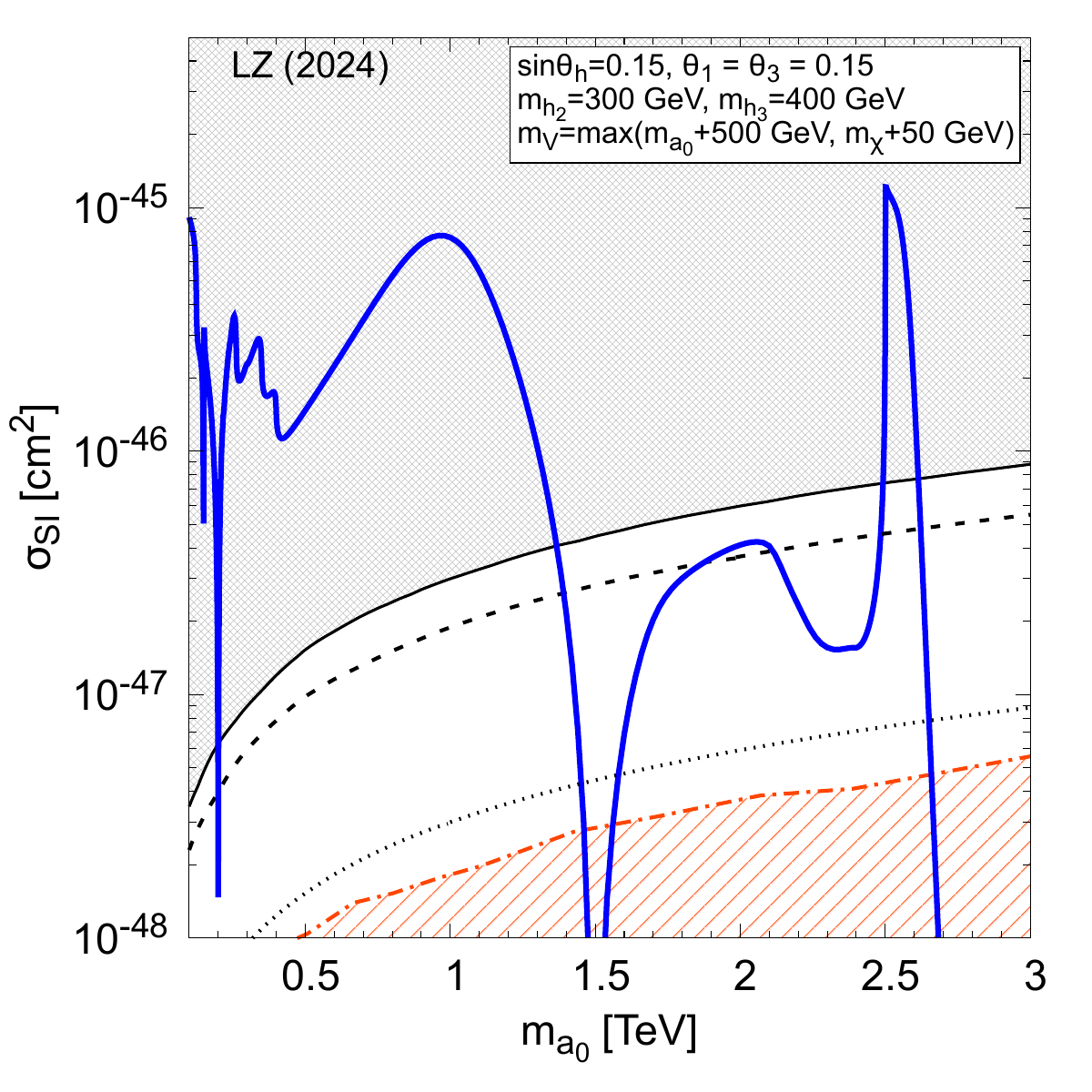}
\includegraphics[width=0.47\hsize]{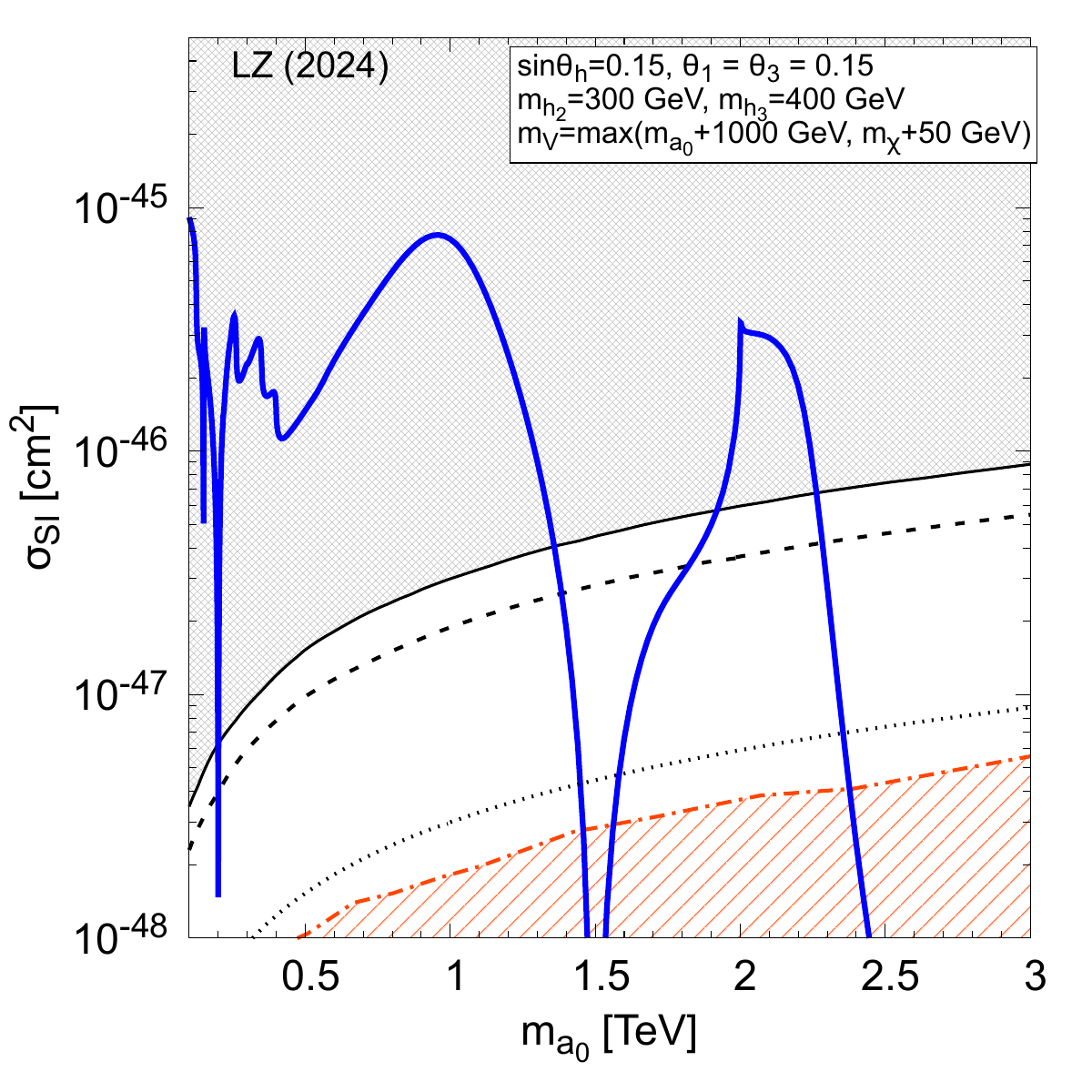}
\caption{
The spin-independent cross section in the $\chi a_0$ scenario.
Here, we use the same parameter values as those in figures~\ref{fig:vs-and-omegah2-ratio_a0_mchi1500} and \ref{fig:omegah2-ratio_a0_mchi1500}. 
The blue-solid curves show $\sigma_\text{SI}$. 
Other color notation is the same as in figure~\ref{fig:chi-V0_sigmaSI_300_400_015}.
}
\label{fig:sigmaSI_a0_mchi1500}
\end{figure}

Figure~\ref{fig:sigmaSI_a0_mchi1500} shows the values of $\sigma_\text{SI}$. Here,
we take the same values of the parameters in figures~\ref{fig:vs-and-omegah2-ratio_a0_mchi1500} and 
\ref{fig:omegah2-ratio_a0_mchi1500}.
For $m_{a_0} \lesssim 0.8 m_\chi$, $\sigma_\text{SI}$ is larger than the current upper bound obtained by the LZ experiment. The exceptions are for $m_{a_0} = m_{h_j}/2$ where a pair of $a_0$ particles annihilates resonantly, and thus the $a_0$ couplings to $h_j$ are tiny to reproduce the right amount of DM energy density. 
For $m_{a_0} \gtrsim m_\chi$, $\sigma_\text{SI}$ can be larger than the neutrino fog and depends on $m_{a_0}$ because the relic abundance of $a_0$ is determined by the bouncing, which is affected by $m_{a_0}$. 
If $m_\chi \simeq m_{V}/2$, $\sigma_\text{SI}$ is even larger than the current upper bound because the number density of $a_0$ is not suppressed for $m_\chi \simeq m_{V}/2$ as discussed in section~\ref{sec:chi-a0_relic}.
For $m_{a_0} \simeq m_\chi$, we find that $\sigma_\text{SI}$ vanishes. 
It is easily understood from eq.~\eqref{eq:Caq}. 
The Wilson coefficient of the effective $a_0$-$q$ interaction is proportional to $m_{a_0}^2 - m_\chi^2$, and thus $\sigma_\text{SI}$ is proportional to the square of it. As a result, $\sigma_\text{SI}$ is highly suppressed for $m_{a_0} \simeq m_\chi$ although the number density of $a_0$ is comparable to that of the pNG DM.
We can also understand this result as a consequence of the enhancement of the global symmetry in the scalar potential in the $m_{a_0} = m_\chi$ limit, which makes $a_0$ pNG as well as $\chi$.
If $m_{a_0} = m_\chi$, then $\lambda_3 = \lambda_4$, and the global symmetry of the scalar quartic terms in the scalar potential is enhanced from $SU(2)_x \times SU(2)_g$ into $SO(4) \times SO(4)$. We discuss the detail of the symmetry structure in appendix~\ref{sec:enhanced-symmetry}.

\begin{figure}[tbp]
\centering
\includegraphics[width=0.47\hsize]{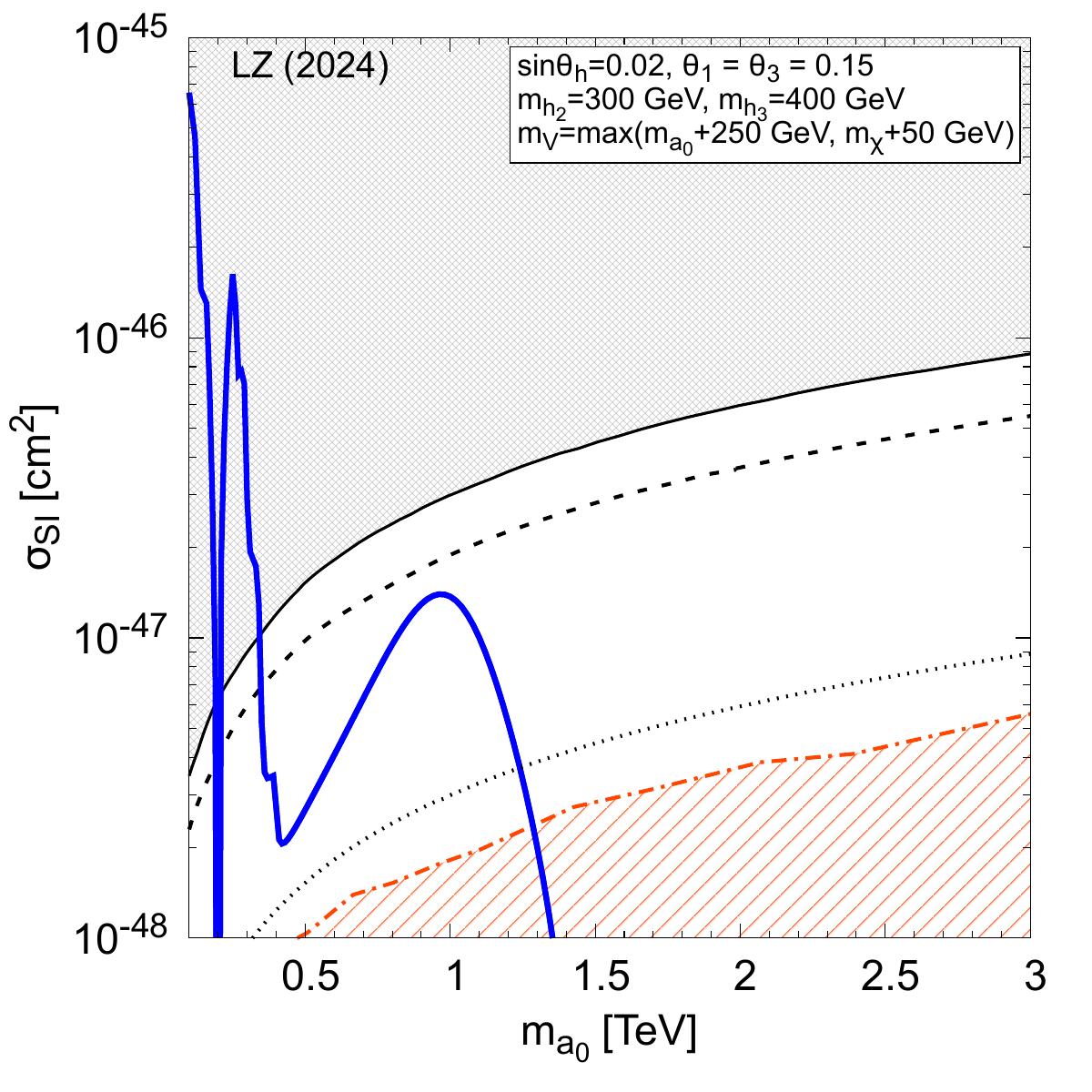}
\caption{
Similar to figure~\ref{fig:sigmaSI_a0_mchi1500} but with smaller $\theta_h$.
}
\label{fig:sigmaSI_a0_mchi1500_light}
\end{figure}

It is also possible to make $\sigma_\text{SI}$ smaller than the current upper bound but larger than the neutrino fog for $m_{a_0} < m_\chi$ by taking smaller $\theta_h$. 
Figure~\ref{fig:sigmaSI_a0_mchi1500_light} shows the case for $\sin\theta_h = 0.02$. 
We find $\sigma_\text{SI}$ is below the current upper bound for $m_{h_3} \lesssim m_{a_0}$.
In contrast to figure~\ref{fig:sigmaSI_a0_mchi1500}, $\sigma_\text{SI}$ is much smaller than neutrino fog for $m_{a_0} > m_\chi$.

\section{Conclusion}\label{sec:conclusion}
We have proposed a new pNG DM model based on a gauged $SU(2)_x$ and a softly broken global $SU(2)_g$ symmetries. These symmetries are spontaneously broken to a $U(1)_D$ global symmetry by the VEV of a scalar field. As a result, a complex pNG boson $\chi$ arises. If $\chi$ is lighter than the $SU(2)_x$ gauge bosons, it is stable due to the $U(1)_D$ global symmetry and is a DM candidate. 
Similar to other pNG DM models, the $\chi$-nucleon elastic scattering cross section relevant for DM direct detection experiments is highly suppressed by the small momentum transfer, whereas the annihilation cross section of a pair of DM particles relevant for the relic abundance is not suppressed. As a result, the model explains the current null results in the direct detection experiments while keeping the thermal DM scenario. 
On the other hand, compared to other pNG DM models, there are several distinguishing features in our model. 
In contrast to the simplest pNG DM model, our model has no global discrete symmetry spontaneously broken, and thus is free from the domain-wall problem. 
Unlike the other pNG models with a $U(1)$ gauge symmetry, the newly introduced gauge coupling in our model does not blow up at higher energy scale because the $SU(2)_x$ gauge symmetry is asymptotic free. 
Also, we do not need an extremely larger energy scale than the DM mass in contrast to some other pNG models with a $U(1)$ gauge symmetry.

Moreover, assuming all the couplings in the scalar potential are real, the model is invariant under the charge conjugation $C_\text{dark}$ in the dark sector and provides two-component DM scenarios depending on the mass spectra. 
We have studied three scenarios:  
(i) the single component scenario in which only the pNG DM is the DM candidate,
(ii) the two-component scenario in which $\chi$ and $V^0$ are the DM candidates,
and 
(iii) the two-component scenario in which $\chi$ and $a_0$ are the DM candidates.

The single component scenario is realized for $m_\chi < 2 m_V$ and $m_\chi < 2 m_{a_0}$. 
In this case, the larger value of $v_s$, which is equivalent to smaller $\chi\chi h_j$ couplings, is required for the heavier $\chi$. This is because a pair of pNG particles mainly annihilates by exchanging $V^\pm$ in the $t$-channel, which depends on the gauge coupling, and thus the small $\chi \chi h_j$ couplings are necessary to avoid too large annihilation cross section. This is a different feature from most of the pNG DM models. 

In the two-component scenarios, we have examined the thermal relic by solving the Boltzmann equation numerically and found that the pNG DM is the dominant component. The other component, $V^0$ or $a_0$, is subdominant in the large regions of the parameter space. 
For $m_V \gtrsim 1.2 m_\chi$, $V^0$ is produced via $\chi \bar{\chi} \to V^0 h_j$, and the time evolution of its yield $Y$ shows the bouncing behavior, as shown in figures \ref{fig:chi-V0_evolution} and \ref{fig:x-vs-Y}. As a result, the energy density of $V^0$ occupies about $\mathcal{O}(1)$\% for $1.2 m_\chi \lesssim m_V \lesssim 2 m_\chi - m_{h_3}$.  
Similarly, $a_0$ is produced via $\chi \bar{\chi} \to a_0 h_j$ for $m_V \gtrsim 1.2 m_\chi$ and is subdominant component. Exception is for $m_\chi \simeq m_V/2$, where the pNG DM resonantly annihilates by exchanging $V^0$ in the $s$-channel, and $a_0$ is the dominant component.

In contrast to the pNG DM, $V^0$ and $a_0$ scatter off a nucleon. However, since they are the subdominant components, the event rates in the direct detection experiments are suppressed by the small number density. The rescaled $V^0$-nucleon and $a_0$-nucleon scattering cross sections according to their number densities are smaller than the current upper bounds in wide parameter regions,
as shown in figures \ref{fig:chi-V0_sigmaSI_300_400_015}, \ref{fig:sigmaSI_a0_mchi1500}, and \ref{fig:sigmaSI_a0_mchi1500_light}. Also, they are above the neutrino fog in the large regions of the parameter space. Therefore, our model gives rise to observable signals in the future direct detection experiments.

\acknowledgments
The authors thank Katsuya Hashino for a collaboration at the early stage.
We draw Feynman diagrams using \texttt{feynMF}~\cite{Ohl:1995kr}.
TA thank Alejandro Ibarra and Stefan Lederer for useful comments.
This work is supported by JSPS Core-to-Core Program (grant number: JPJSCCA20200002) and JSPS KAKENHI Grant Numbers 21K03549.

\appendix
\section{Some identities for constructing scalar potential}\label{app:some-identities}

We define $\tilde{\Phi}$ by
\begin{align}
\tilde{\Phi} 
\equiv 
- \epsilon \Phi^* \epsilon
=
\begin{pmatrix}
 \tilde{\varphi}_2 & \varphi_1
\end{pmatrix}
.
\end{align}
Under the $SU(2)_x \times SU(2)_g$, $\tilde{\Phi}$ transforms as the same as $\Phi$,
$\tilde{\Phi} \to U_x \tilde{\Phi} U_g^\dagger$.
We find $\tilde{\Phi}^\dagger$ is related to $\Phi$ as
\begin{align}
 - \epsilon \Phi^t \epsilon = \tilde{\Phi}^\dagger.
\end{align}
Using $\tilde{\Phi}$, we find the following identities.
\begin{align}
 \det(\Phi) =& \frac{1}{2} \tr(\tilde{\Phi}^\dagger \Phi),\\
 \det(\tilde{\Phi}) =& \frac{1}{2} \tr(\tilde{\Phi}^\dagger \tilde{\Phi}) = \left(\det(\Phi)\right)^\dagger,\\
 \det(\tilde{\Phi}^\dagger \Phi \tau_3) =& 0,\\
 \det(\Phi^\dagger \tilde{\Phi} \tau_3) =& 0,\\
 \tr(\Phi^\dagger \Phi \Phi^\dagger \Phi) =& \left( \tr(\Phi^\dagger \Phi)\right)^2 - 2 \det(\tilde{\Phi})\det(\Phi),\\
 \tr(\tilde{\Phi^\dagger}\tilde{\Phi}\tilde{\Phi^\dagger}\tilde{\Phi}) =&  \tr(\Phi^\dagger \Phi \Phi^\dagger \Phi) ,\\
 \tr(\tilde{\Phi}^\dagger \Phi \Phi^\dagger \Phi) =&  \det(\Phi)\tr(\Phi^\dagger \Phi) ,\\
 \tr(\Phi^\dagger \tilde{\Phi} \Phi^\dagger \Phi) =&  \det(\tilde{\Phi})\tr(\Phi^\dagger \Phi) ,\\
 \tr(\tilde{\Phi^\dagger}\tilde{\Phi}\tilde{\Phi^\dagger} \Phi) =&  \tr(\tilde{\Phi}^\dagger \Phi \Phi^\dagger \Phi) ,\\
 \tr(\tilde{\Phi^\dagger}\tilde{\Phi} \Phi^\dagger\tilde{\Phi}) =&  \tr(\Phi^\dagger \tilde{\Phi} \Phi^\dagger \Phi) ,\\
 \tr(\tilde{\Phi^\dagger}\tilde{\Phi} \Phi^\dagger \Phi) =&  -  \tr(\Phi^\dagger \Phi \Phi^\dagger \Phi) + \left( \tr(\Phi^\dagger \Phi)\right)^2,\\
 \tr(\tilde{\Phi^\dagger} \Phi \tilde{\Phi}^\dagger \Phi) =&  2 \left( \det(\Phi) \right)^2,\\
 \tr(\tilde{\Phi^\dagger} \Phi \Phi^\dagger \tilde{\Phi}) =&  -  \tr(\Phi^\dagger \Phi \Phi^\dagger \Phi) + \left( \tr(\Phi^\dagger \Phi)\right)^2,\\
 \tr(\Phi^\dagger \tilde{\Phi} \Phi \tilde{\Phi}) =&  2 \left( \det(\tilde{\Phi}) \right)^2.
\end{align}
These identities are useful to construct the scalar potential.

\section{Accidental symmetry for the degenerate gauge boson mass}\label{app:custodial}
As briefly discussed in section~\ref{sec:gauge-sector}, the model has an approximate global $SU(2)$ symmetry that results in the degenerate mass of the $SU(2)_x$ gauge bosons.
In this section, we discuss the symmetry more explicitly.
The symmetry rotating the would-be NG bosons ($\pi_{V^+}, \pi_{V^0}, \pi_{V^-}$) is a global $SU(2)$ symmetry.
It is useful to express scalar fields by the following two-by-two matrices, 
\begin{align}
\Phi_1 = \mqty(\tilde{\varphi}_1, \varphi_1), \quad
\Phi_2 = \mqty(\tilde{\varphi}_2, \varphi_2),
\end{align}
where $\tilde{\varphi}_j = \epsilon \varphi_j^*$. 
Using these matrices, it is easy to find that 
\begin{align}
\varphi_1^\dagger \varphi_1 =& \frac{1}{2} \tr(\Phi_1^\dagger \Phi_1),\\
\varphi_2^\dagger \varphi_2 =& \frac{1}{2} \tr(\Phi_2^\dagger \Phi_2),\\
\varphi_1^\dagger \varphi_2 + \varphi_2^\dagger \varphi_1 =& \tr(\Phi_2^\dagger \Phi_1),\\
\varphi_1^\dagger \varphi_2 - \varphi_2^\dagger \varphi_1 =& \tr(\Phi_2^\dagger \tau^3 \Phi_1).
\end{align}
Under the global $SU(2)$ symmetry, $\Phi_1$, $\Phi_2$, and $V_\mu$ transforms as
\begin{align}
 \Phi_1 \to& U \Phi_1 U^\dagger,\\
 \Phi_2 \to& U \Phi_2 U^\dagger,\\
 V_\mu \to& U V_\mu U^\dagger,
\end{align}
where $U$ is a two-by-two matrix of the global $SU(2)$ symmetry.
All the other fields remain unchanged under this global $SU(2)$ transformation. 
In this transformation, $\tr(\Phi_1^\dagger \Phi_1)$, $\tr(\Phi_2^\dagger \Phi_2)$, $\tr(\Phi_2^\dagger \Phi_1)$, 
and $D^\mu \varphi_1^\dagger D_\mu \varphi_1^\dagger + D^\mu \varphi_2^\dagger D_\mu \varphi_2 = \frac{1}{2} \tr(D^\mu \Phi_1^\dagger D_\mu \Phi_1) + \frac{1}{2} \tr(D^\mu \Phi_2^\dagger D_\mu \Phi_2) $ 
 are invariant, 
but $\tr(\Phi_2^\dagger \tau^3 \Phi_1)$ is not.
Under the assumption that all the couplings in the scalar potential are real, which is taken through in this paper, only a linear combination of the $\lambda_3$ and $\lambda_4$ terms contains $\tr(\Phi_2^\dagger \tau^3 \Phi_1)$,
\begin{align}
\lambda_3 \qty((\varphi_1^\dagger \varphi_2)^2 + (\varphi_2^\dagger \varphi_1)^2) + 2 \lambda_4 (\varphi_1^\dagger \varphi_2) (\varphi_2^\dagger \varphi_1) 
=& \frac{\lambda_3 + \lambda_4}{2} \tr(\Phi_2^\dagger \Phi_1) + \frac{\lambda_3 - \lambda_4}{2} \tr(\Phi_2^\dagger \tau^3 \Phi_1).
\end{align}
Therefore, a finite value of $\lambda_3 - \lambda_4$ is the source of the global $SU(2)$ symmetry breaking and generates the mass difference between $V^\pm$ and $V^0$ at the loop level.
Using eqs.~\eqref{eq:lam3} and \eqref{eq:lam4}, we find $\lambda_3 - \lambda_4 = (m_\chi^2 - m_{a_0}^2)/v_s^2$. Thus, $m_{V^+} \neq m_{V^0}$ at the loop level if $m_\chi = m_{a_0}$.

At the one-loop level, we find 
\begin{align}
 m_{V^+}^2 - m_{V^0}^2
=& 
\frac{g_D^2}{(4\pi)^2}
\sum_j R_{1j}^2 \qty(  B_{00}(m_{V}, m_\chi^2, m_{h_j}^2) - B_{00}(m_{V}, m_{a_0}^2, m_{h_j}^2))
\nonumber\\
&+
\frac{g_D^2}{(4\pi)^2} \qty( B_{00}(m_{V}^2, m_{a_0}^2, m_\chi^2) - B_{00}(m_{V}^2, m_\chi^2, m_\chi^2) )
,
\end{align}
where $B_{00}$ is a loop function given by \texttt{LoopTools}~\cite{Hahn:1998yk}.

\section{Global symmetry in the scalar potential}\label{sec:enhanced-symmetry}
We discuss the symmetry structure of the scalar potential eq.~\eqref{eq:potential} in the $\lambda_3 = \lambda_4$ limit, discussed in section~\ref{sec:DD_chi-a}.
For that purpose, we express the component fields of $\varphi_1$ and $\varphi_2$ as
\begin{align}
 \varphi_1 =& \mqty( \phi_3 + i \phi_4 \\ \phi_1 + i \phi_2),\ \
 \varphi_2 = \mqty( \rho_3 + i \rho_4 \\ \rho_1 + i \rho_2),
\end{align}
where $\phi_j$ and $\rho_j$ are real scalar fields, and also introduce
\begin{align}
 \vb*{\phi} = \mqty(\phi_1 \\ \phi_2 \\ \phi_3 \\ \phi_4), \ \ 
 \vb*{\rho} = \mqty(\rho_1 \\ \rho_2 \\ \rho_3 \\ \rho_4).
\end{align}
The scalar potential consists of the following combinations of the fields,
\begin{align}
 \mathcal{O}_1 =&
 \varphi_1^\dagger \varphi_1 + \varphi_2^\dagger \varphi_2
= \sum_{j=1}^4 \qty( \phi_j^2 + \rho_j^2 )
= \mqty(\vb*{\phi}^t & \vb*{\rho}^t) \mqty(\vb*{\phi} \\ \vb*{\rho}),
\\
 \mathcal{O}_2 =&
 \varphi_1^\dagger \varphi_1 - \varphi_2^\dagger \varphi_2
= \mqty(\vb*{\phi}^t & \vb*{\rho}^t) \mqty(1 & 0 \\ 0 & -1) \mqty(\vb*{\phi} \\ \vb*{\rho}),
\\
 \mathcal{O}_+ =&
 \varphi_1^\dagger \varphi_2 + \varphi_2^\dagger \varphi_1
= 2 \sum_{j=1}^4  \phi_j \rho_j
= \mqty(\vb*{\phi}^t & \vb*{\rho}^t) \mqty(0 & 1 \\ 1 & 0) \mqty(\vb*{\phi} \\ \vb*{\rho}),
\\
 \mathcal{O}_- =&
 \varphi_1^\dagger \varphi_2 - \varphi_2^\dagger \varphi_1
= 2 i \qty( \phi_1 \rho_2 - \phi_2 \phi_1 + \phi_3 \phi_4 - \phi_4 \phi_3  )
= \mqty(\vb*{\phi}^t & \vb*{\rho}^t) \mqty(0 & E \\ -E & 0) \mqty(\vb*{\phi} \\ \vb*{\rho})
,
\end{align}
where
\begin{align}
 E =& \mqty( 0 & 1 & 0 & 0\\ -1 & 0 & 0 & 0 \\ 0 & 0 & 0 & 1 \\ 0 & 0 & -1 & 0 ).
\end{align} 
With these notation, the scalar potential eq.~\eqref{eq:potential} is expressed as
\begin{align}
 V_0 + V_\text{soft} 
=& \mu_1^2 \mathcal{O}_1 + \mu_2^2 \mathcal{O}_+ 
+ \frac{\mu_\chi^2}{2} \mathcal{O}_2
+ \tilde{\lambda}_1 H^\dagger H \mathcal{O}_1
+ \tilde{\lambda}_2 H^\dagger H \mathcal{O}_1
\nonumber\\
&
+ \frac{\lambda_1}{2} (\mathcal{O}_1)^2
+ \lambda_2 \mathcal{O}_1 \mathcal{O}_+
+ \frac{\lambda_3 + \lambda_4}{2} (\mathcal{O}_+)^2
+ \frac{\lambda_3 - \lambda_4}{2} (\mathcal{O}_-)^2
\nonumber\\
&
+ \mu_H^2 H^\dagger H + \frac{\lambda_H}{2}(H^\dagger H)^2
.
\end{align}
In the $\lambda_3 = \lambda_4$ limit, the potential is independent from $\mathcal{O}_-$.
Also, $\mathcal{O}_2$ appears only in the $\mu_\chi^2$ term, which is the soft symmetry breaking term. 
To understand the symmetry of the potential, we need to discuss the symmetry of each $\mathcal{O}_i$.

It is apparent that $\mathcal{O}_1$ has $O(8)$ symmetry. We focus on its continuous part $SO(8)$ here. Under the $SO(8)$ transformation, $\mqty(\vb*{\phi}^t & \vb*{\rho}^t )^t$ transforms as
\begin{align}
 \mqty(\vb*{\phi} \\ \vb*{\rho}) 
\to U \mqty(\vb*{\phi} \\ \vb*{\rho}),
\end{align}
where $U$ is a $8\times 8$ matrix for the $SO(8)$ transformation.
For the infinitesimal transformation, $U$ is given by
\begin{align}
 U \simeq 1 + \mqty( A & C \\ -C^t & B),
\end{align}
where $A$, $B$, and $C$ are $4 \times 4$ matrices and satisfy $A^t = -A$ and $B^t = - B$.
The other $\mathcal{O}_i$ give constraints on $A$, $B$, and $C$, and break the $SO(8)$ symmetry.

Under the infinitesimal $SO(8)$ transformation, $\mathcal{O}_+$ transforms as
\begin{align}
\mqty(\vb*{\phi}^t & \vb*{\rho}^t) \mqty(0 & 1 \\ 1 & 0) \mqty(\vb*{\phi} \\ \vb*{\rho})
\to&
 \mqty(\vb*{\phi}^t & \vb*{\rho}^t) U^t \mqty(0 & 1 \\ 1 & 0) U\mqty(\vb*{\phi} \\ \vb*{\rho})
\nonumber\\
\simeq&
 \mqty(\vb*{\phi}^t & \vb*{\rho}^t) 
\left\{
 \mqty(0 & 1 \\ 1 & 0)
+ \mqty( -C - C^t & B-A \\ A-B & C+C^t)
\right\}
\mqty(\vb*{\phi} \\ \vb*{\rho})
.
\end{align}
Therefore, $\mathcal{O}_+$ breaks the $SO(8)$ symmetry, but is invariant if $A=B$ and $C=-C^t$.
Namely, $\mathcal{O}_+$ is invariant under $SO(4) \times SO(4)$.

Under the infinitesimal $SO(8)$ transformation, $\mathcal{O}_2$ transforms as
\begin{align}
\mqty(\vb*{\phi}^t & \vb*{\rho}^t) \mqty(1 & 0 \\ 0 & -1) \mqty(\vb*{\phi} \\ \vb*{\rho})
\to&
 \mqty(\vb*{\phi}^t & \vb*{\rho}^t) U^t \mqty(1 & 0 \\ 0 & -1) U\mqty(\vb*{\phi} \\ \vb*{\rho})
\nonumber\\
\simeq&
 \mqty(\vb*{\phi}^t & \vb*{\rho}^t) 
\left\{
 \mqty(1 & 0 \\ 0 & -1) 
+ \mqty( 0 & 2C \\ 2C^t & 0)
\right\}
\mqty(\vb*{\phi} \\ \vb*{\rho})
.
\end{align}
Therefore, $\mathcal{O}_2$ is not invariant under the $SO(8)$ but is invariant if $C=0$.

We count the number of (pseudo) NG bosons in the $\lambda_3 = \lambda_4$ limit.
If we ignore the soft symmetry breaking term, the potential consists of $\mathcal{O}_1$ and $\mathcal{O}_+$ and is invariant under the $SO(4) \times SO(4)$ global symmetry,
\begin{align}
 U \simeq 1 + \mqty( A & C \\ C & A),
\end{align}
where $A$ and $C$ are $4\times 4$ antisymmetric matrices.
The VEV in eq.~\eqref{eq:desired-vacuum} breaks this symmetry spontaneously into
$SO(3) \times SO(3)$. This spontaneous symmetry breaking generates six NG bosons.
Adding the soft symmetry breaking term, $\mathcal{O}_2$ modifies the symmetry structure of the potential, and the potential is not invariant under $SO(4) \times SO(4)$ but is $SO(4)$,
\begin{align}
 U \simeq 1 + \mqty( A & 0 \\ 0 & A).
\end{align}
This symmetry is spontaneously broken into $SO(3)$ by the VEV. 
Therefore, the number of true NG bosons is three, and the rest are pseudo-NG bosons, which consists of $\chi$, $\bar{\chi}$, and $a_0$.

\bibliographystyle{JHEP} 
\bibliography{refs.bib}

\end{document}